\documentclass[12pt]{article}
\usepackage[utf8]{inputenc} 
\usepackage[binary-units]{siunitx}
\usepackage{cite}
\usepackage{graphicx}
\usepackage{tikz}
\usepackage{url}
\usepackage[a4paper, total={7in, 8.5in}]{geometry}
\usepackage{pgfplots}
\usepackage{amsmath}
\usepackage{amssymb}
\usepackage{hyperref}
\usepackage{xcolor}

\pgfplotsset{compat=1.17}
\begin{document}
\title{Large, ultra-flat optical traps for uniform quantum gases}
\date{}
\author{Kai Frye-Arndt $^{1,2}$*,  Matthew Glaysher$^1$, Marius Glaeser$^1$, \\ Matthias Koch$^1$, Stefan Seckmeyer$^1$, Holger Ahlers$^{2}$, Waldemar Herr$^{2}$,\\ Naceur Gaaloul$^1$, Christian Schubert$^{2}$ and Ernst Maria Rasel$^1$ \\
$^1$ Institute for Quantenoptik,\\ Leibniz University Hannover, 30167 Hannover\\ 
$^2$ German Aerospace Center (DLR),\\ Institute for Satellite Geodesy and Inertial Sensing,\\ 30167 Hannover, Germany\\
* frye@iqo.uni-hannover.de}
\maketitle
\begin{abstract}
Ultracold atomic gases with uniform density can be created by flat-bottom optical traps.
These gases provide an ideal platform to study many-body physics in a system that allows for simple connections with theoretical models and emulation of numerous effects from a wide range of fields of physics.
In Earth-bound laboratories the trap sizes, number of species and states, as well as the range of physical effects are largely restricted by the adopted levitation technique. Homogeneous ultracold gases in microgravity simulators and space however offer an interesting perspective which is actively being pursued.
To exploit the full potential of any gravity-compensated laboratory the box potentials created need to be as large as possible. 
By using two orthogonally aligned acousto-optic deflectors, we create large time-averaged optical potentials with trapping volumes a thousandfold larger than conventional setups, described by power-law scalings with exponents of up to $152$. 
We verify the performance of our setup by simulating the mean-field behaviour of a quantum gas ground state in conjunction with dynamical excitations due to the realistic time-dependent painting potentials. 
The implementation of this setup may open new directions at the interface with condensed matter, few-body Efimov physics or the exploration of critical, non-equilibrium phenomena.

\vspace{2pc}

\end{abstract}

\section{Introduction}\label{sec:Introduction}
Blue-detuned light exerts a repulsive force on atoms
that can be used to confine them into a dark, central spot surrounded by repulsive barriers.
Such traps have been used to create uniform Bose-Einstein condensates~\cite{Gaunt2013}, observe critical dynamics of spontaneous symmetry breaking~\cite{Navon2015}, study superfluid transition in a spin-balanced Fermi gas~\cite{Mukherjee2017} and explore many more phenomena~\cite{Navon2021}.
Since microgravity eliminates the need for levitation techniques, it provides a uniquely clean environment for studying ultracold quantum gases.
On Earth, gravity necessitates levitation methods using either electromagnetic fields, e.g. dc electric fields in case of polar molecules~\cite{Bause2021}, dc magnetic fields~\cite{Gaunt2013}, or ac fields in the radio, microwave and optical frequency range.
However, all of these techniques are prone to spatial inhomogeneties, impose stringent requirements on temporal stability and feature numerous restrictions. 
In fact, levitation techniques inevitably distort the quantum gases, which primarily emerge from technical noise, such as fluctuations in the levitation fields or imperfections in their spatial and temporal stability, or unwanted deviations from an isotropic trapping potential.
Additionally, the compensation of gravitational pull is inherently dependent on the mass and internal state of the atoms being manipulated. 
Different atomic species and their quantum states require tailored compensation strategies to counteract gravity effectively. 

For these reasons, levitation techniques constrain the trapping volume to typically $\SI{100}{\micro\meter}\times\SI{100}{\micro\meter}\times\SI{50}{\micro\meter}$~\cite{Gaunt2013,Schymik2022,Hilker2022,Mukherjee2017}.
In the case of magnetic levitation, Maxwell's equation \mbox{$\nabla \cdot \Vec{B} = 0$} compromises homogeneity of the trap bottom and leads to horizontal magnetic field gradients for cylindrically symmetric configurations~\cite{Li2015}.
Regarding optical levitation~\cite{Shibata2020}, large traps require high laser powers, which makes generating flat-bottom potentials a challenging task even at the depth scale of the nanokelvin.
Larger flat traps increase the sensitivity to critical dynamics at larger length scales~\cite{Navon2015} and enhance the sharpness of the trap edges.
A parametrization for this sharpness is the exponent $p$ of a power-law potential $V(r) \sim r^p$ -- the larger the exponent $p$, the more box-like the potential. 
State-of-the-art experiments achieve values of $p \approx 10$, yet beyond-mean-field physics, e.g. modification of the critical temperature of interacting Bose gases, is only revealed for $p \gtrsim 100$~\cite{Navon2015,Zobay2005}.

The restrictions related to levitation can be lifted by operating in microgravity conditions.
Large flat-bottom traps in microgravity would enable the investigation of a plethora of novel effects ranging from thermodynamic properties, phase transitions, hydrodynamics of spinor gases, miscibility of mixtures, etc. (see ref.~\cite{Stamper-Kurn2023} for an extensive overview).
Furthermore, control over the trapping geometry in combination with the long coherence times expected in blue-detuned traps~\cite{Friedmann2000} would allow for the creation of quantum information storage~\cite{Jessen2004}, e.g. via coupling to a microwave resonator~\cite{Xiang2013}.
Although blue-detuned painted potentials are particularly well-suited for microgravity applications, their full potential for compact mobile setups has yet to be demonstrated. 
Unlocking this capability would advance quantum-enhanced sensors for navigation, geodesy, climate change research, and fundamental physics~\cite{Acin2018, Knight2019, Raymer2019}.

Here we introduce a setup for creating large, blue-detuned optical traps for ultracold atoms with an exceptionally dark trap center.
We present our compact and robust mechanical setup in section~\ref{sec:Design}.
We also showcase the versatility of the creation as well as characterize the performance of optical traps (section~\ref{sec:Realizing}) and determine the residual scattering rate and optical potential in the trap center to show the remarkable flatness of the trap bottom.
Because our setup surpasses previous optical trap sizes by over an order of magnitude per dimension, we are able realize power-law traps with exponents reaching up to 152.
We perform numerical simulations (section~\ref{sec:theory}) to show the feasibility of trapping Bose-Einstein condensates in blue-detuned time-averaged traps.
Finally, we discuss our results (section~\ref{sec:Discussion}).

\section{Results}

\subsection{Experimental design and setup}\label{sec:Design}
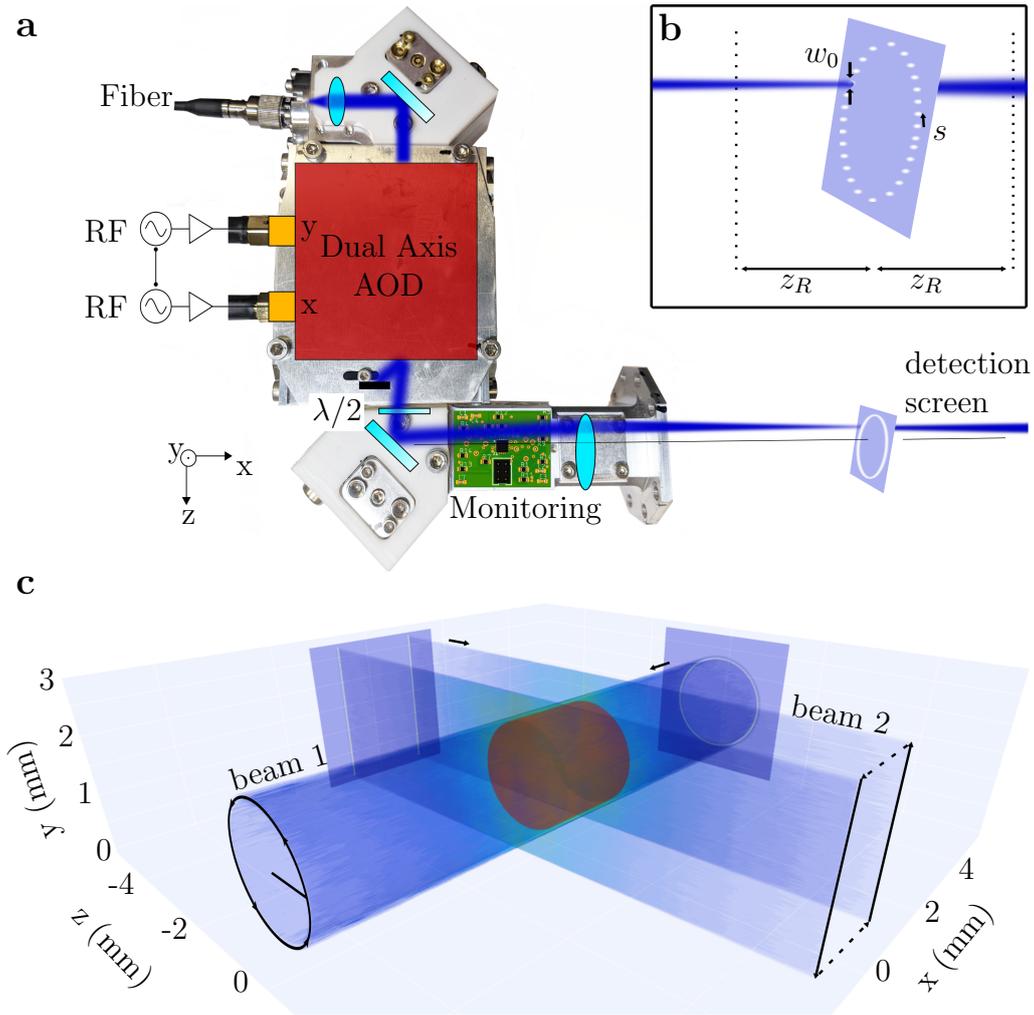
\begin{figure}[ht!]
\centering
	  \begin{tikzpicture}[>=stealth]
	\node[inner sep=0pt] (Picture) at (0,0)
	{\includegraphics[width=370.5pt]{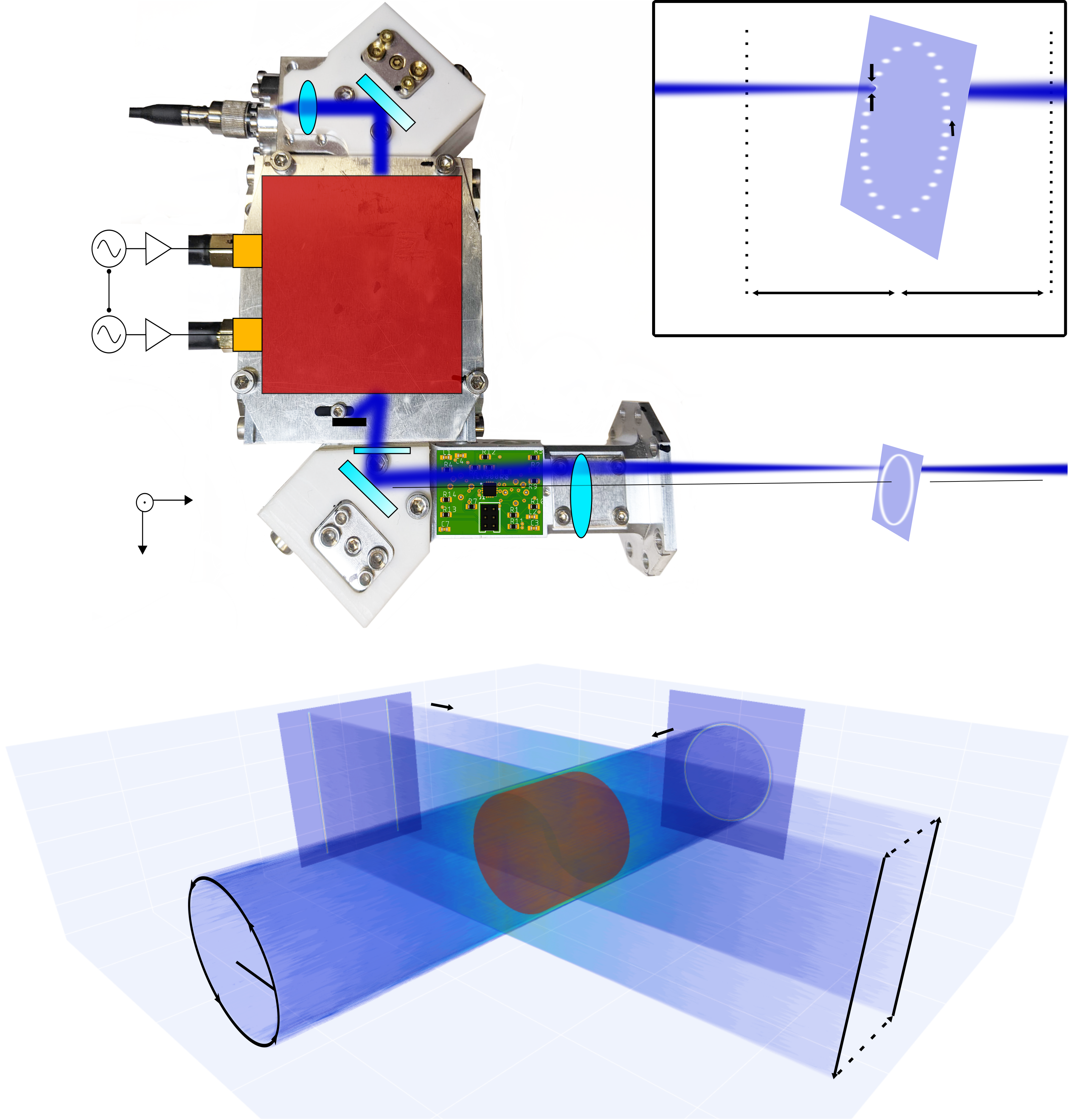}};

	\node[anchor=west] at (-7.2,6.4) {\large \textbf{a}};
	\node[anchor=west] at (1.25,6.4) {\large \textbf{b}};
	\node[anchor=west] at (-7.2,-1) {\large \textbf{c}};
	
	\def\rend{5.4}
	\def\expr{-3.6}
	\def\lend{-5.4}

    \node[anchor=west] at (-6.3,3.7) {RF};
    \node[anchor=west] at (-6.3,2.7) {RF};
    \node[anchor=west] at (-6.1,5.5) {Fiber};

    \node[anchor=west] at (-3.45,3.7) {y};
    \node[anchor=west] at (-3.45,2.7) {x};
    
    \node[anchor=west, align=center] at (-3.2, 3.25) {Dual Axis\\ AOD };
    
    \node[fill=white, inner sep=1pt, anchor=west] at (-3.2, 1.3) {$\lambda/2$};

    
    \node[anchor=west] at (-5.2,0.75) {y};
    \node[anchor=west] at (-4.3,0.55) {x};
    \node[anchor=west] at (-5.02,-0.1) {z};
    
    \node[anchor=west] at (-1.5,0) {Monitoring};


    \node[anchor=center] at (3.21,3) {$z_R$};
    \node[anchor=center] at (4.92,3) {$z_R$};
    \node[anchor=center] at (5.1,5) {$s$};
    \node[anchor=center] at (3.55,6) {$w_0$};
    \node[anchor=west, align=left] at (4.5,1.7) {detection\\ screen};

    \node[anchor=center, rotate=16] at (-3.6, -3.4) {beam 1};
    \node[anchor=center, rotate=-11] at (3.8, -2.7) {beam 2};
	\def\yhigh{0.75}
	\def\yleft{0.25}
	\def\ybottom{-4.5}
	\def\yll{-5.65}
	
	\node[anchor=center, rotate=108] at (-6.8, -3.7) {y (mm)};
	
	\node[anchor=east] at (\yll - 0*\yleft, \ybottom + 0*\yhigh) {0};
	\node[anchor=east] at (\yll - 1*\yleft, \ybottom + 1*\yhigh) {1};
	\node[anchor=east] at (\yll - 2*\yleft, \ybottom + 2*\yhigh) {2};
	\node[anchor=east] at (\yll - 3*\yleft, \ybottom + 3*\yhigh) {3};
	
	\def\zhigh{0.8}
	\def\zslope{0.9}
	\def\zbottom{-6.25}
	\def\zll{-3.85}
	
	\node[anchor=center, rotate=-46] at (-5.8,-5.8) {z (mm)};
	
	\node[anchor=east] at (\zll - 0*\zslope, \zbottom + 0*\zhigh) {0};
	\node[anchor=east] at (\zll - 1*\zslope+0.1, \zbottom + 1*\zhigh-0.1) {-2};
	\node[anchor=east] at (\zll - 2*\zslope+0.3, \zbottom + 2*\zhigh-0.3) {-4};

	\def\xhigh{0.77}
	\def\xslope{-0.65}
	\def\xbottom{-6.1}
	\def\xll{4.60}
	
	\node[anchor=center, rotate=51] at (5.3, -5.8) {x (mm)};
	
	\node[anchor=east] at (\xll - 0*\xslope, \xbottom + 0*\xhigh) {0};
	\node[anchor=east] at (\xll - 1*\xslope, \xbottom + 1*\xhigh) {2};
	\node[anchor=east] at (\xll - 2*\xslope-0.2, \xbottom + 2*\xhigh-0.2) {4};
	
	\end{tikzpicture}
  \caption{Concept for implementing time-averaged potentials. \textbf{a} A dual-axis AOD, controlled by phase-coupled RF sources, deflects a laser beam in x- and y-directions. A lens converts angles to parallel displacement and focuses the beam. A detection screen is placed at the lens's focal length. \textbf{b} A zoomed view of the detection screen shows the beam divergence characterized by the Rayleigh range $z_R$ and waist $w_0$. The AOD displaces the beam in steps of size $s$. \textbf{c} Volumetric intensity representation of a 3D trapping potential using experimental data. Images are recorded every \SI{250}{\micro\meter} as the detection screen moves axially. The traps from beam 1 and beam 2 are combined via software, forming a red cylindrical trapping volume.}\label{fig:Painting_Potentials}
\end{figure}

The centerpiece of our setup is a dual-axis acousto-optic deflector (AOD, DTSXY-400 from \textit{AAOptoelectronics}) which dynamically steers a laser beam to generate versatile optical potentials (see figure~\ref{fig:Painting_Potentials}(a)), a process often referred to as "painting".
Painted potentials have found extensive applications in ultracold atom research, including ring traps~\cite{Henderson2009, Bell2016} that underpin atomtronics~\cite{Amico2021}, optical tweezers for individual atom and molecule control~\cite{Kaufman2021}, collimation of matter waves~\cite{Albers2022}, gravity compensation~\cite{Shibata2020} and accelerated optical evaporation~\cite{herbst2022}. 
However, we are not aware of their application for creating blue-detuned traps until now.

AODs were chosen to realize box traps due to their exceptional technical advantages~\cite{Gauthier2021}, even though they only create optical traps in a time-averaged regime.
While spatial light modulators (SLM)~\cite{Gaunt2013} or digital mirror device (DMDs)~\cite{Amico2021}, are alternatives for generating static traps, they require bulky, power-intensive controllers for pixel-level manipulation of the light field's amplitude or phase.
In contrast, AODs are driven by easily generated radio-frequency (RF) signals, offer deflection efficiencies exceeding \SI{80}{\percent}, and require less optical path length for separating diffraction orders.
Additionally, AODs do not require frequent phase-delay recalibration, which would otherwise be necessary due to temperature variations.
Consequently, AODs are particularly suitable for compact and remote-use setups.
Our compact design with a footprint of $\SI{10.5}{\centi \meter} \times \SI{5.8}{\centi \meter} \times \SI{14.4}{\centi \meter}$ and a weight of just \SI{477}{\g} was successfully tested against vibrations typical of launch conditions for space missions, such as those to the ISS~\cite{NanoracksIDD2020} and sounding rockets~\cite{Grosse2016}. 
This robustness qualifies it for space-borne experiments~\cite{Frye2021}.

\begin{figure}[ht]
\centering
	  \begin{tikzpicture}[>=stealth]
	\node[inner sep=0pt] (Picture) at (0,0)
	{\includegraphics[width=351pt]{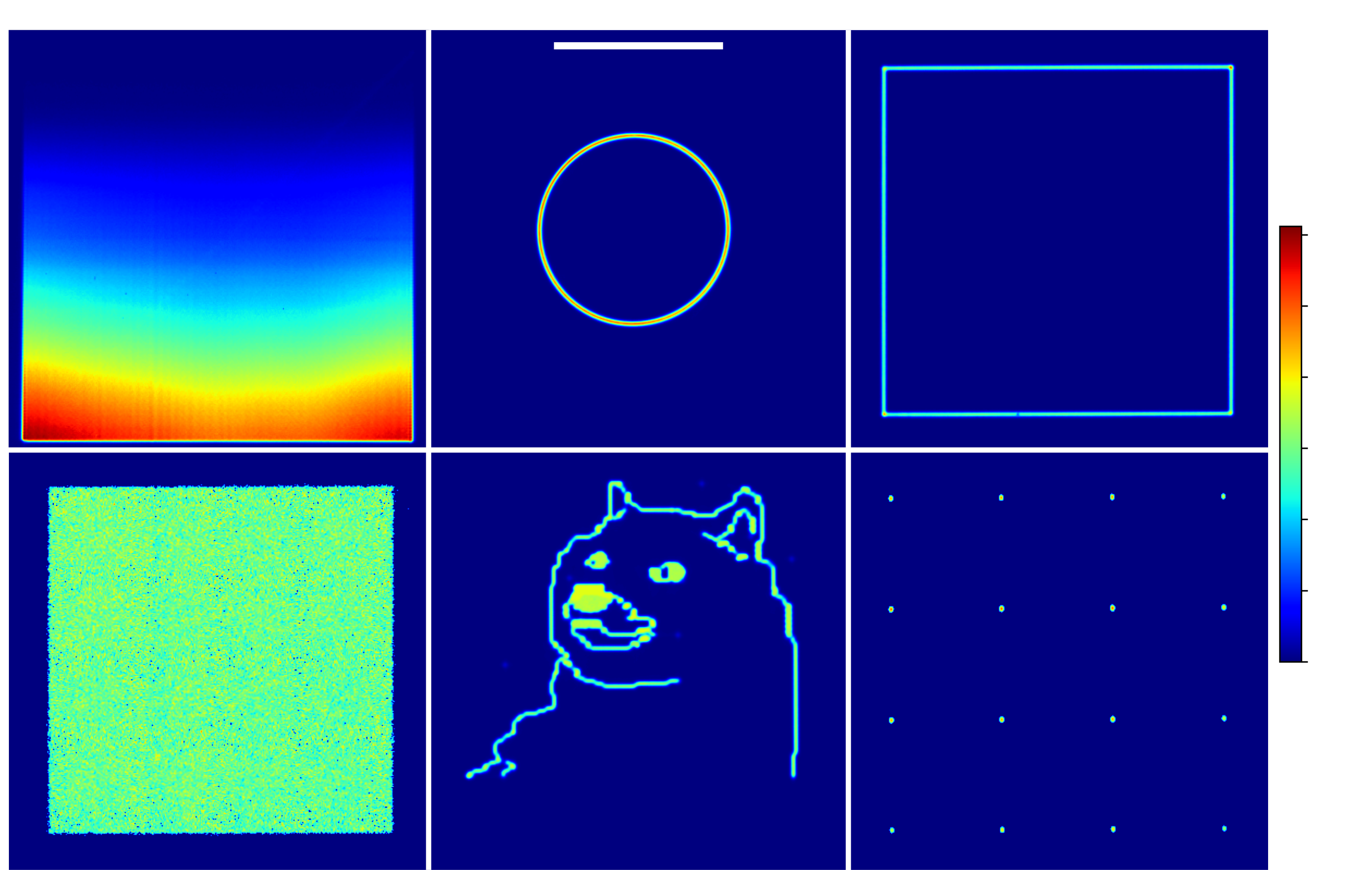}};
	
	\def\rend{5.4}
	\def\expr{-3.6}
	\def\lend{-5.4}
	\node[anchor=center, color=white] at (-0.5,3.28) {\SI{1}{\milli\meter}};
	\node[anchor=west] at (5.25,2.2) {$\mathrm{I_{max}}$};
	\node[anchor=west] at (5.28,-2.3) {0};

    \def\x{-5.9}
    \def\xadd{3.8}
    \def\y{3.45}
    \def\yadd{-3.88}

    \node[anchor=center, color=white] at (\x + 0*\xadd, \y + 0*\yadd) {\large \textbf{a}};
    \node[anchor=center, color=white] at (\x + 1*\xadd, \y + 0*\yadd) {\large \textbf{b}};
    \node[anchor=center, color=white] at (\x + 2*\xadd, \y + 0*\yadd) {\large \textbf{c}};
    \node[anchor=center, color=white] at (\x + 0*\xadd, \y + 1*\yadd) {\large \textbf{d}};
    \node[anchor=center, color=white] at (\x + 1*\xadd, \y + 1*\yadd) {\large \textbf{e}};
    \node[anchor=center, color=white] at (\x + 2*\xadd, \y + 1*\yadd) {\large \textbf{f}};
	\end{tikzpicture}
  \caption{Six exemplary optical potentials representing \textbf{a} an intensity gradient with curvature, \textbf{b} a circle, \textbf{c} a square box, \textbf{d} an equally illuminated area, \textbf{e} the outline of a Shiba Inu dog and \textbf{f} a $4\times4$ array of dots. The camera is fixed in the focal plane. The size of all images is indicated in the center of the top row.}\label{fig:Potentials_exmaples}
\end{figure}

The working principle is as follows (figure~\ref{fig:Painting_Potentials}(a)):
The light is injected into the setup via a polarization maintaining optical fiber and diffracted by the AOD with efficiencies of $>$\SI{80}{\percent} at the central frequency.
The control electronics updates the RF signal every $\Delta \tau= \SI{2.64}{\micro\second}$ (see section \ref{Sec:Electronics}) matching the AOD's optical response time of \SI{2}{\micro \second}.
The AOD device is manufactured such that the doubly deflected beam is nearly collinear to the incident beam.
The optical system after the AOD is adjustable to correct for imperfections in the collinearity, and a wave plate ($\lambda/2$ in figure~\ref{fig:Painting_Potentials}) in the beam path allows for fine control of the beam polarization, which is crucial for avoiding spurious lattices in 3D trap configurations.
For remote applications where direct access to the setup is not possible, such as on the ISS~\cite{Frye2021} or sounding rockets~\cite{Becker2018}, monitoring is implemented in the setup to collect additional data relevant for interpreting experimental results and housekeeping.
Partially reflected light is used for monitoring the intensity, polarization and position of the light beam, using two duo-lateral position sensitive diodes (PSD).
Finally, a lens with $f=\SI{80}{\milli\meter}$ translates the angle of the deflected beam after the AOD into a parallel displacement and focuses the beam onto the atom plane with a waist of $w_0=\SI{19}{\micro\meter}$.

Our setup operates at an optical wavelength of \SI{764}{\nano\meter}, making it ideal for manipulating widely used elements like potassium and rubidium.
To create 3D traps two of the AOD setups need to be combined, preferably orthogonal to each other.
For instance, intersecting a circularly moving beam with two light sheets creates a closed cylinder as shown in figure~\ref{fig:Painting_Potentials}(b).
We focus on a single AOD setup, since the second setup can be constructed identically to create a 3D potential.
The expansion of the beam around the focus leads to a non-uniform trap depth in axial direction.
However, the axial extension of the trap is below the Rayleigh range and the beam expands only by a factor of 1.375.
This has marginal effect on the trapping volume, but the trap depth is reduced by a factor of two. 

Figure~\ref{fig:Potentials_exmaples} demonstrates the versatility of the setup in generating optical potentials, showcasing its ability to control optical power, create both connected and separated shapes, e.g. to realize multiple traps, as well as fill large areas. 
Our AOD setup can paint shapes within a square area of up to $\left(\SI{2.8}{\milli\meter}\right)^2$, which is more than an order of magnitude larger per dimension than most current implementations.

\subsection{Realizing a "blue box" for atoms}\label{sec:Realizing}
In this section, we evaluate key features of the created uniform trap, such as the atom-light scattering rate, the flatness of the trap center, and the sharpness of the power-law potential.

\begin{figure}[t]
\centering
	  \begin{tikzpicture}[>=stealth]
	\node[inner sep=0pt] (Picture) at (0,0)
	{\includegraphics[width=351pt]{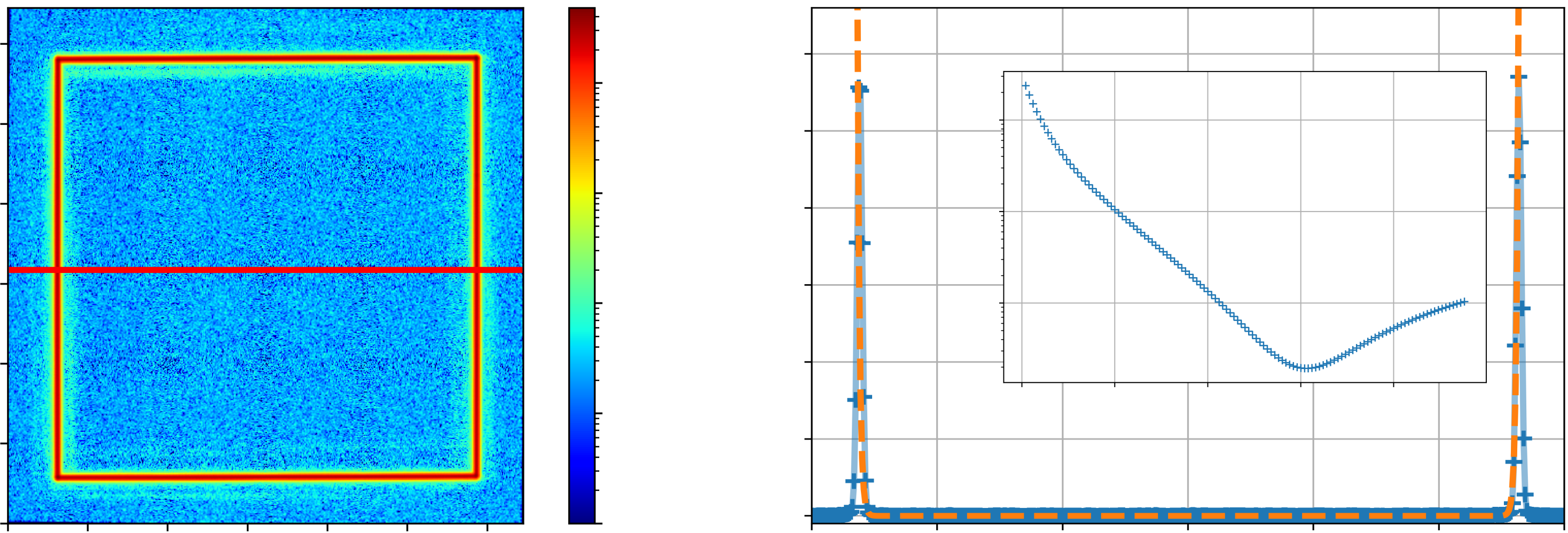}};

	\node[anchor=center] at (-6.3,2.45)  {\large \textbf{a}};
 	\node[anchor=center] at (-0,2.45)  {\large \textbf{b}};
	\node[rotate=90] at (-7.1, 0) {y (mm)};
	
	\def\spacingRx{0.64}
	\def\Lupperleft{1.8}
	\def\rgr{-6.1}
	\node[anchor=east] at (\rgr,\Lupperleft-0*\spacingRx) {3};
	\node[anchor=east] at (\rgr,\Lupperleft-1*\spacingRx) {2.5};
	\node[anchor=east] at (\rgr,\Lupperleft-2*\spacingRx) {2};
	\node[anchor=east] at (\rgr,\Lupperleft-3*\spacingRx) {1.5};
	\node[anchor=east] at (\rgr,\Lupperleft-4*\spacingRx) {1};
	\node[anchor=east] at (\rgr,\Lupperleft-5*\spacingRx) {0.5};
	\node[anchor=east] at (\rgr,\Lupperleft-6*\spacingRx) {0};

	\node[anchor=center] at (-4,-2.8) {x (mm)};
	\def\bgr{-2.28}
	\def\spacingLx{0.62}
    \def\Llowerleft{-6.1}	
	\node[anchor=center] at (\Llowerleft-0*\spacingLx,\bgr) {0};
	\node[anchor=center] at (\Llowerleft+1*\spacingLx,\bgr) {0.5};
	\node[anchor=center] at (\Llowerleft+2*\spacingLx,\bgr) {1};
	\node[anchor=center] at (\Llowerleft+3*\spacingLx,\bgr) {1.5};
	\node[anchor=center] at (\Llowerleft+4*\spacingLx,\bgr) {2};
	\node[anchor=center] at (\Llowerleft+5*\spacingLx,\bgr) {2.5};
	\node[anchor=center] at (\Llowerleft+6*\spacingLx,\bgr) {3};

    \def\low{-2.}
    \def\spacing{0.87}
    \def\xvalue{-1.5}
    \node[anchor=west] at (\xvalue,\low+0*\spacing) {$10^0$};
    \node[anchor=west] at (\xvalue,\low+1*\spacing) {$10^1$};
    \node[anchor=west] at (\xvalue,\low+2*\spacing) {$10^2$};
    \node[anchor=west] at (\xvalue,\low+3*\spacing) {$10^3$};
    \node[anchor=west] at (\xvalue,\low+4*\spacing) {$10^4$};
    \node[anchor=west] at (-1.5, 2.2) {$I_{max}$};

	\node[rotate=90] at (-0.2, 0.1) {intensity (arb.)};
	\def\spacingLx{0.995}
    \def\Llowerleft{0.2}	
    \node[anchor=center] at (3.5,-2.8) {x (mm)};
	\node[anchor=center] at (\Llowerleft+0*\spacingLx,\bgr) {-1.5};
	\node[anchor=center] at (\Llowerleft+1*\spacingLx,\bgr) {-1.0};
	\node[anchor=center] at (\Llowerleft+2*\spacingLx,\bgr) {-0.5};
	\node[anchor=center] at (\Llowerleft+3*\spacingLx,\bgr) {0};
	\node[anchor=center] at (\Llowerleft+4*\spacingLx,\bgr) {0.5};
	\node[anchor=center] at (\Llowerleft+5*\spacingLx,\bgr) {1.0};
	\node[anchor=center] at (\Llowerleft+6*\spacingLx,\bgr) {1.5};

	\node[rotate=90] at (0.85, 0.1) {residuals $\bar{r}$};
	\def\bgr{-1.15}
	\def\spacingLx{0.72}
    \def\Llowerleft{1.91}	
	\node[anchor=center] at (\Llowerleft+0*\spacingLx,\bgr) {0};
	\node[anchor=center] at (\Llowerleft+1*\spacingLx,\bgr) {50};
	\node[anchor=center] at (\Llowerleft+2*\spacingLx,\bgr) {100};
	\node[anchor=center] at (\Llowerleft+3*\spacingLx,\bgr) {150};
	\node[anchor=center] at (\Llowerleft+4*\spacingLx,\bgr) {200};

	\node[anchor=center] at (3.5,-1.6) {$p$};
	
	\def\xpos{1.43}
	\def\lowery{-0.92}
	\def\spacing{0.68}
	\node[anchor=center] at (\xpos,\lowery+0*\spacing) {$10^4$};
	\node[anchor=center] at (\xpos,\lowery+1*\spacing) {$10^5$};
	\node[anchor=center] at (\xpos,\lowery+2*\spacing) {$10^6$};
	\node[anchor=center] at (\xpos,\lowery+3*\spacing) {$10^7$};

	\end{tikzpicture}

  \caption{Realization of a large box potential with ultra-flat trap bottom. \textbf{a} We show a square shaped optical potential. Note that the intensity scaling is logarithmic showcasing the dark trap center. The red line indicates the 1D slice through the center that we show in \textbf{b}. \textbf{b} Optimisation of a power law $x^p$, shown by the orange dotted line, to the measured intensity values (blue crosses). See section~\ref{sec:AppendixExponent} details on the fitting routine. The inset shows the squared mean residuals for different power law exponents $p$. The residual is minimal for an exponent $p=152$.}\label{fig:LargeBox}
\end{figure}
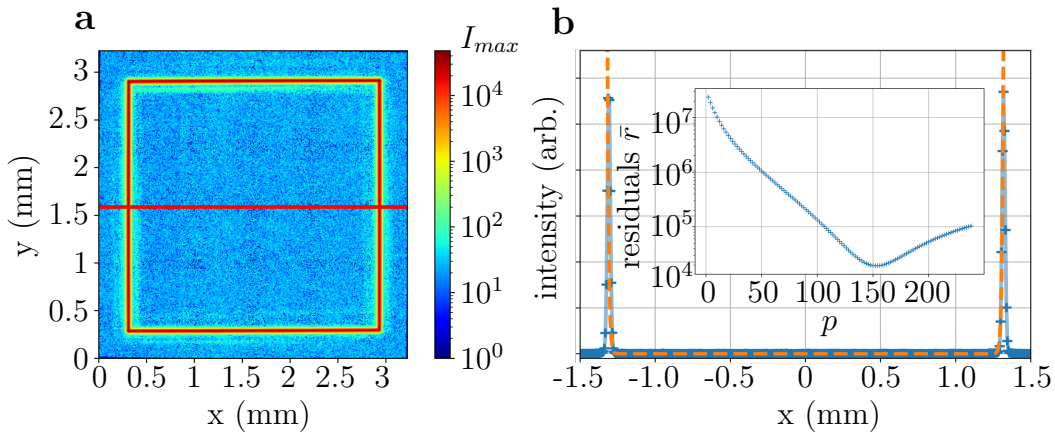

The scattering rate is a critical parameter as it directly impacts the lifetime of the condensate inside the trap by inducing heating and losses. It is determined by the residual light intensity at the trap center~\cite{Grimm2000}.
Figure~\ref{fig:LargeBox}(a) displays a CCD image of a quadratically shaped potential with logarithmic intensity scaling demonstrating a dark trap center.
Due to the limited dynamic range of the camera, we used an alternative approach to measure residual intensity: a pinhole with a diameter of \SI{0.7}{\milli\meter} was placed at the center of a painted circle of \SI{2.8}{\milli\meter} diameter and we recorded the light power that passes the hole.
We find that this power corresponds to $\num{1.3e-4}$ of the optical power along the circle resulting in an upper bound for the scattering rate per optical power $P$ of the trapping beam of $\Gamma_{sc}/P=2\pi\times\SI{2.08}{\milli\hertz\per W}$ for $^{87}$Rb. 
Considering that only a few \SI{}{\milli W} of optical power are sufficient to sustain atom trapping (see section~\ref{sec:theory}), we conclude that losses due to atom-light scattering are therefore negligible compared to other losses, such as background vacuum pressure.

Similarly, we find the residual optical potential inside the trap center to be \SI{0.14}{\nano K \per W}.
This represents an improvement of four orders compared to the SLM-based trap reported in ref.~\cite{Martirosyan2024}, underscoring the exceptional flatness of the trap bottom.

Finally, we estimate the sharpness $p$ of the created optical potentials, which depends on the size of the trap.
A trap diameter smaller than $\approx \SI{50}{\micro\meter}$ yields a power-law trap with $p=2$, which corresponds to a harmonic trap.
For larger traps the center becomes progressively flatter, while confinement is provided primarily at the edges.
Figure~\ref{fig:LargeBox}(b) illustrates a large quadratically shaped potential, where the flat center and confinement at the edges are evident.
Fitting a power law to this box to determine the sharpness is non-trivial, since the potential is mostly consistent with zero and only a few pixels provide signals above zero.
We developed a fitting method (detailed in section~\ref{sec:AppendixExponent}) and found that $p=152$ minimizes the mean squared residual.
This sharpness is an order of magnitude higher than what is previously reported~\cite{Navon2021}.

\subsection{Bose-Einstein condensates in the "blue box" potential}\label{sec:theory}
Time-averaged optical potentials have proven their usefulness in creating red-detuned traps~\cite{herbst2022,Albers2022,Bell2016} but to our knowledge the blue-detuned counterpart was not yet demonstrated. 
We use numerical simulations to verify the ability of our system to create box-shaped BECs.
The uniformity of the resulting atomic densities, as well as the required painting frequency to justify a time-averaged description are investigated. 
Finally, we show the feasibility of our system to create trap diameters of $\approx$ \SI{2.8}{\milli\meter}.

The system is described by the Gross-Pitaevskii equation (GPE)~\cite{Pethick_Smith2008} (see equation~\ref{equ:GPE} in section~\ref{sec:Reduction}), where the shape of the entered time-dependent potential $V(\textbf{r},t)$ governs the dynamics of the resulting wave functions. 
The equation is solved numerically via the Split Operator Method~\cite{Feit_1982}. 
Throughout this section, we assume the 3D box potentials to be cylindrical in shape, where the height of the cylinder $L_z$ is small compared to the radial extent.
We begin by employing the imaginary time propagation~\cite{Bao_2004} to calculate ground states, which requires static potentials. 
Therefore an average of all spatially separated Gaussian beams along the circumference is taken.
The resulting ground state of the system is then propagated with real time painting to study the influence that the dynamically changing potentials have on the atoms. For reference, to create the dynamics shown in the discussion in figure \ref{fig:2D_energies} the position of the Gaussian beam is moved on the order of every $\SI{0.1}{\milli\second}$ to $\SI{10}{\milli\second}$, depending on the painting frequency, while the response of the atoms is numerically resolved at $\approx \SI{26}{\micro\second}$.

A stable trap configuration is obtained by finding a balance between the density and the potential trapping energy, where the interactions are large enough to create the box-like shapes, but not too large as to push out the atoms from the trap.
When expanding to large trap sizes this balance remains, provided the density is kept constant by tuning the atom numbers. 

Solving the GPE in 3D is computationally challenging, especially for the exploration of large box potentials.
We therefore perform a reduction to an effective 2D system (section ~\ref{sec:Reduction}) and verify the accuracy of the assumptions by a comparison to a full 3D calculation for a single smaller scale realization.
The ground states for a true 3D system and the effective 2D system show that our reduction method leads to valid results, allowing us to only perform 2D calculations in the following section.

\subsubsection{Ground states in static traps}\label{sec:StaticPotential}
\begin{figure}[ht]
\centering
	  \begin{tikzpicture}[>=stealth]
	\node[inner sep=0pt] (Picture) at (0,0)
	{\includegraphics[width=382.2pt]{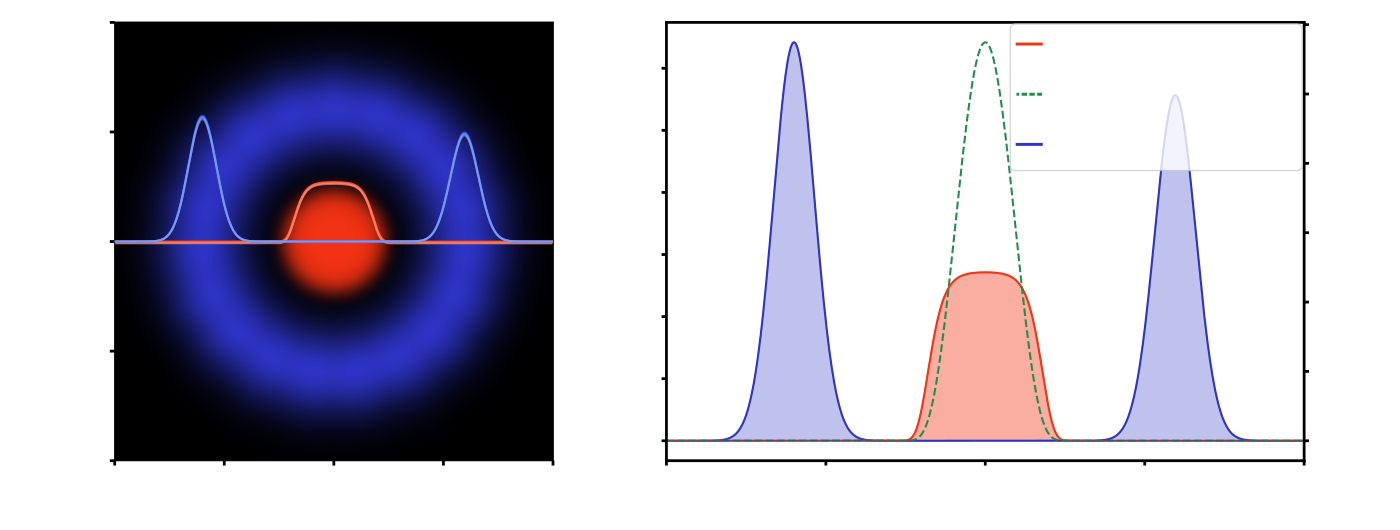}};

    \node[anchor=center, color=white] at (-5.15, 2.0) {\textbf{a}};
    \node[anchor=center] at (0.2, 2.0) {\textbf{b}};
    \node[anchor=center] at (-3.45,-2.5) {x (µm)};
    \def\y{-2.1}
    \def\xstart{-5.58}
    \def\xstep{1.03}
    \node[anchor=center] at (\xstart + 0*\xstep,\y) {-200};
    \node[anchor=center] at (\xstart + 1*\xstep,\y) {-100};
    \node[anchor=center] at (\xstart + 2*\xstep,\y) {0};
    \node[anchor=center] at (\xstart + 3*\xstep,\y) {100};
    \node[anchor=center] at (\xstart + 4*\xstep,\y) {200};

    \node[anchor=center, rotate=90] at (-6.5,0.5) {y (µm)};
    \def\x{-5.6}
    \def\ystart{-1.8}
    \def\ystep{1.03}
    \node[anchor=east] at (\x, \ystart + 0*\ystep) {-200};
    \node[anchor=east] at (\x, \ystart + 1*\ystep) {-100};
    \node[anchor=east] at (\x, \ystart + 2*\ystep) {0};
    \node[anchor=east] at (\x, \ystart + 3*\ystep) {100};
    \node[anchor=east] at (\x, \ystart + 4*\ystep) {200};

    \node[anchor=center] at (2.9,-2.5) {x (µm)};
    \def\y{-2.1}
    \def\xstart{-0.15}
    \def\xstep{1.45}
    \node[anchor=center] at (\xstart + 0*\xstep,\y) {-200};
    \node[anchor=center] at (\xstart + 1*\xstep,\y) {-100};
    \node[anchor=center] at (\xstart + 2*\xstep,\y) {0};
    \node[anchor=center] at (\xstart + 3*\xstep,\y) {100};
    \node[anchor=center] at (\xstart + 4*\xstep,\y) {200};

    \node[anchor=center, rotate=90, color=red] at (-1,0.3) {2D $|\Psi|^2$ ($10^{-4}$/µm$^2$)};
    \def\x{-0.3}
    \def\ystart{-1.65}
    \def\ystep{1.19}
    \node[anchor=east, color=red] at (\x, \ystart + 0*\ystep) {0};
    \node[anchor=east, color=red] at (\x, \ystart + 1*\ystep) {2};
    \node[anchor=east, color=red] at (\x, \ystart + 2*\ystep) {4};
    \node[anchor=east, color=red] at (\x, \ystart + 3*\ystep) {6};

    \node[anchor=center, rotate=90, color=blue] at (6.8,0.3) {$\mathrm{E_{pot}}$ (µK)};
    \def\x{6.6}
    \def\ystart{-1.65}
    \def\ystep{1.3}
    \node[anchor=east, color=blue] at (\x, \ystart + 0*\ystep) {0.0};
    \node[anchor=east, color=blue] at (\x, \ystart + 1*\ystep) {0.1};
    \node[anchor=east, color=blue] at (\x, \ystart + 2*\ystep) {0.2};
    \node[anchor=east, color=blue] at (\x, \ystart + 3*\ystep) {0.3};

    \def\x{3.2}
    \def\ystart{2.05}
    \def\ystep{0.48}
    \node[anchor=west] at (\x, \ystart - 0*\ystep) {\small{$|\Psi_0|^2,\,$N=7500}};
    \node[anchor=west] at (\x, \ystart - 1*\ystep) {\small{$|\Psi_0|^2,\,$N=0}};
    \node[anchor=west] at (\x, \ystart - 2*\ystep) {\small{$\mathrm{E_{pot}}$}};
    \end{tikzpicture}
  \caption{Simulated box-like BEC occupying the ground state of the painted optical ring potential calculated in 2D. \textbf{a} The x-y-view of the interacting atom cloud density $|\Psi_0|^2$ (red) (equation~\ref{equ:GPE}), trapped by the outer circle of optical potential energy $E_{pot}$ (blue) in arbitrary units as a visual reference. This state appears for $N=7500$~atoms confined in a box with a diameter of $d_{box}=\SI{241.9}{\micro\meter}$ and a vertical extent of $L_z=\SI{20}{\micro\meter}$. The extent $L_z$ serves solely to account for the assumed reduction to an effective interaction strength. The trap consists of 19 Gaussians, radially spaced every $s=\SI{40}{\micro\meter}$ along the circumference, generated using a painting laser with power $P_{pl}=\SI{5}{\milli\watt}$. The radius is chosen such that the number of Gaussians per cycle remains an integer. To determine the ground state, a time-average over all Gaussian beams is applied. The red and blue color intensities represent the density and energy, respectively. The overlaid curves correspond to the slice at $y=0$. \textbf{b} Slice across the box potential in physical units. The dashed green line is the non-interacting solution of the same system ($N=0$ in equation~\ref{equ:GPE}) and serves as a comparison to the BEC. The observed box-like density profile of the atom cloud arises from repulsive atom-atom interactions within a sufficiently sharp potential box.
 }\label{fig:2D_densities}
\end{figure}
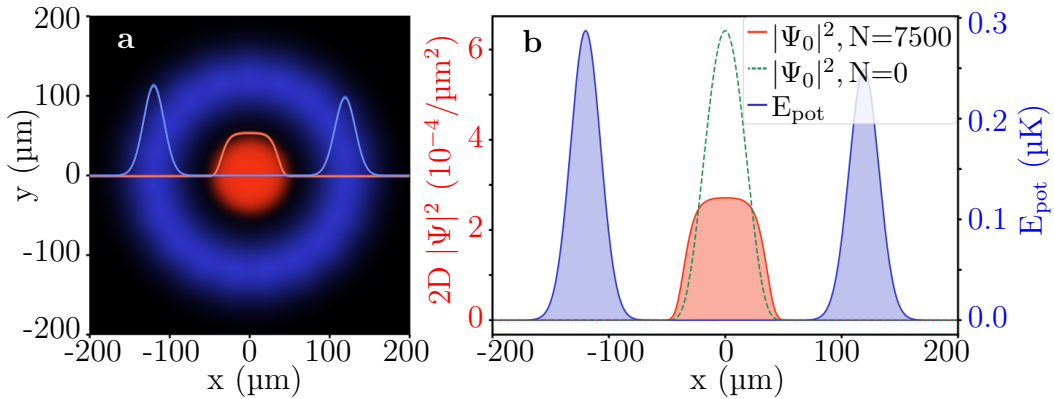

With the methods described in section~\ref{sec:theory} we find the ground state of an effective 2D system with $N=7500$ atoms and a diameter of $d_{box} = \SI{241.9}{\micro\meter}$ (figure~\ref{fig:2D_densities}(a)), which is comparable to other realizations of uniform potentials~\cite{Gaunt2013}.
The atomic density features a flat plateau in the center of the trap and all simulations show negligible atom loss over their run time.  
To emphasize the uniformity of the cloud, we compare in figure~\ref{fig:2D_densities}(b) the ground state of the BEC (red) to the ground state of the non-interacting gas, which is obtained by solving the Schrödinger equation of the same potential (dashed green). 
This shows, that the repulsive atom-atom interactions drive the atom cloud to become box-shaped when faced with a sufficiently sharp potential (blue).

\begin{figure}[ht]
\centering
    \begin{tikzpicture}[>=stealth]
	\node[inner sep=0pt] (Picture) at (0,0)
	{\includegraphics[width=382.2pt]{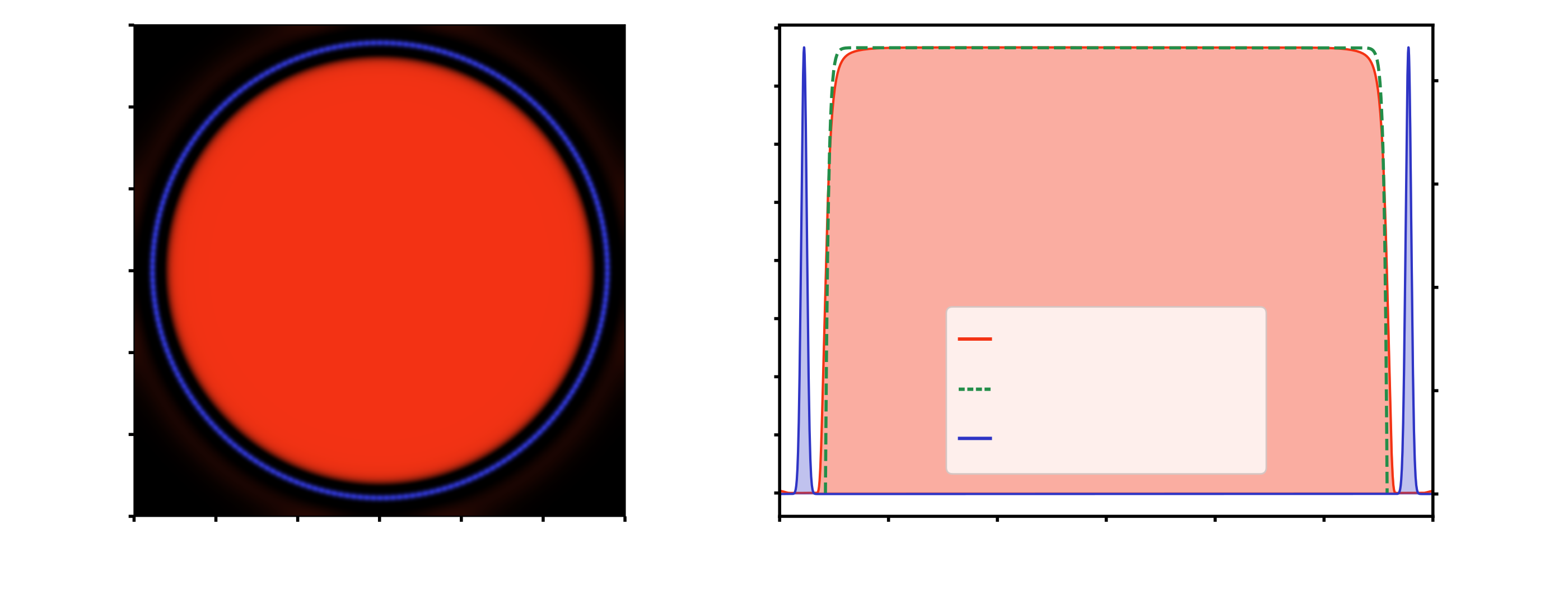}};
    \node[anchor=center, color=white] at (-5.15, 1.8) {\textbf{a}};
    \node[anchor=center] at (0.9, 1.8) {\textbf{b}};
    \node[anchor=center] at (-3.45,-2.5) {x (µm)};
    \def\y{-2.1}
    \def\xstart{-5.58}
    \def\xstep{1.4}
    \node[anchor=center] at (\xstart + 0*\xstep,\y) {-1500};
    \node[anchor=center] at (\xstart + 1*\xstep,\y) {-500};
    \node[anchor=center] at (\xstart + 2*\xstep,\y) {500};
    \node[anchor=center] at (\xstart + 3*\xstep,\y) {1500};

    \node[anchor=center, rotate=90] at (-6.6,0.5) {y (µm)};
    \def\x{-5.55}
    \def\ystart{-1.75}
    \def\ystep{1.34}
    \node[anchor=east] at (\x, \ystart + 0*\ystep) {-1500};
    \node[anchor=east] at (\x, \ystart + 1*\ystep) {-500};
    \node[anchor=east] at (\x, \ystart + 2*\ystep) {500};
    \node[anchor=east] at (\x, \ystart + 3*\ystep) {1500};

    \node[anchor=center] at (2.9,-2.5) {x (µm)};
    \def\y{-2.1}
    \def\xstart{0.2}
    \def\xstep{1.72}
    \node[anchor=center] at (\xstart + 0*\xstep,\y) {-1500};
    \node[anchor=center] at (\xstart + 1*\xstep,\y) {-500};
    \node[anchor=center] at (\xstart + 2*\xstep,\y) {500};
    \node[anchor=center] at (\xstart + 3*\xstep,\y) {1500};

    \node[anchor=center, rotate=90, color=red] at (-0.9,0.3) {2D $|\Psi|^2$ ($10^{-7}$/µm$^2$)};
    \def\x{0.0}
    \def\ystart{-1.65}
    \def\ystep{1.0}
    \node[anchor=east, color=red] at (\x, \ystart + 0*\ystep) {0.0};
    \node[anchor=east, color=red] at (\x, \ystart + 1*\ystep) {0.5};
    \node[anchor=east, color=red] at (\x, \ystart + 2*\ystep) {1.0};
    \node[anchor=east, color=red] at (\x, \ystart + 3*\ystep) {1.5};
    \node[anchor=east, color=red] at (\x, \ystart + 4*\ystep) {2.0};

    \node[anchor=center, rotate=90, color=blue] at (6.6,0.3) {$\mathrm{E_{pot}}$ (µK)};
    \def\x{6.3}
    \def\ystart{-1.65}
    \def\ystep{1.3}
    \node[anchor=east, color=blue] at (\x, \ystart + 0*\ystep) {0.0};
    \node[anchor=east, color=blue] at (\x, \ystart + 1*\ystep) {0.1};
    \node[anchor=east, color=blue] at (\x, \ystart + 2*\ystep) {0.2};
    \node[anchor=east, color=blue] at (\x, \ystart + 3*\ystep) {0.3};

    \def\x{1.8}
    \def\ystart{-0.35}
    \def\ystep{0.48}
    \node[anchor=west] at (\x, \ystart - 0*\ystep) {\small{$|\Psi_0|^2,\,$N=$10^6$}};
    \node[anchor=west] at (\x, \ystart - 1*\ystep) {\small{$\sim x^{99.1}$}};
    \node[anchor=west] at (\x, \ystart - 2*\ystep) {\small{$\mathrm{E_{pot}}$}};
    \end{tikzpicture}
  \caption{Simulation of a very large homogeneous box-like BEC ground state in two dimensions.
  \textbf{a} The x-y-view of the interacting atom cloud density $|\Psi_0|^2$ (red) (equation~\ref{equ:GPE}), trapped by the outer circle of optical potential energy in arbitrary units as a visual reference. The circular box potential has a diameter of $d_{box}=\SI{2775.66}{\micro\meter}$, allowing for 218 painted Gaussians along the circumference of the box at $s=\SI{40}{\micro\meter}$ spacing. For the purpose of finding the ground state, a static average of all Gaussians is taken. To be comparable to figure~\ref{fig:2D_densities}, the same laser power of $P_{pl}=\SI{5}{\milli\watt}$ and z-extent $L_z=20\mu m$ were chosen. The atom number was scaled up accordingly to $N=10^6$ to keep the densities and the interaction regime constant.  The intensity of the red and blue are proportional to the density and the energy, respectively. \textbf{b} Slice at $y=\SI{0}{\mu\meter}$ showing the flatness of the atom cloud density. A power law fit $\sim x^{p}$ results in $p = 99.1$. }\label{fig:2D_densities_bigbox}
\end{figure}

Next, we increase the diameter of the box to $d_{box}=\SI{2775.66}{\micro\meter}$ (figure~\ref{fig:2D_densities_bigbox}).
To keep the same density as for the smaller box, the number of atoms is increased to $N=10^6$, which is an experimentally achievable atom number\cite{Rudolph2015,Gaunt2013}. 
The simulations show that the atom-atom interactions of the density were sufficient to create the box-like cloud (red) with the highly homogeneous part covering most of the trap extent (blue). 
It is common practice to fit the power law $\sim x^{p}$ exponent to describe the "boxiness" of the BEC ~\cite{Navon2021}. 
As seen in figure \ref{fig:2D_densities_bigbox}(b) for the green dashed line, in our case such a fit yields a $p = 99.1$.   

\subsubsection{Real time propagation of dynamic painting}\label{sec:Checking}

\begin{figure}[ht]
\centering

    \begin{tikzpicture}[>=stealth]
	\node[inner sep=0pt] (Picture) at (0,0)
	{\includegraphics[width=234pt]{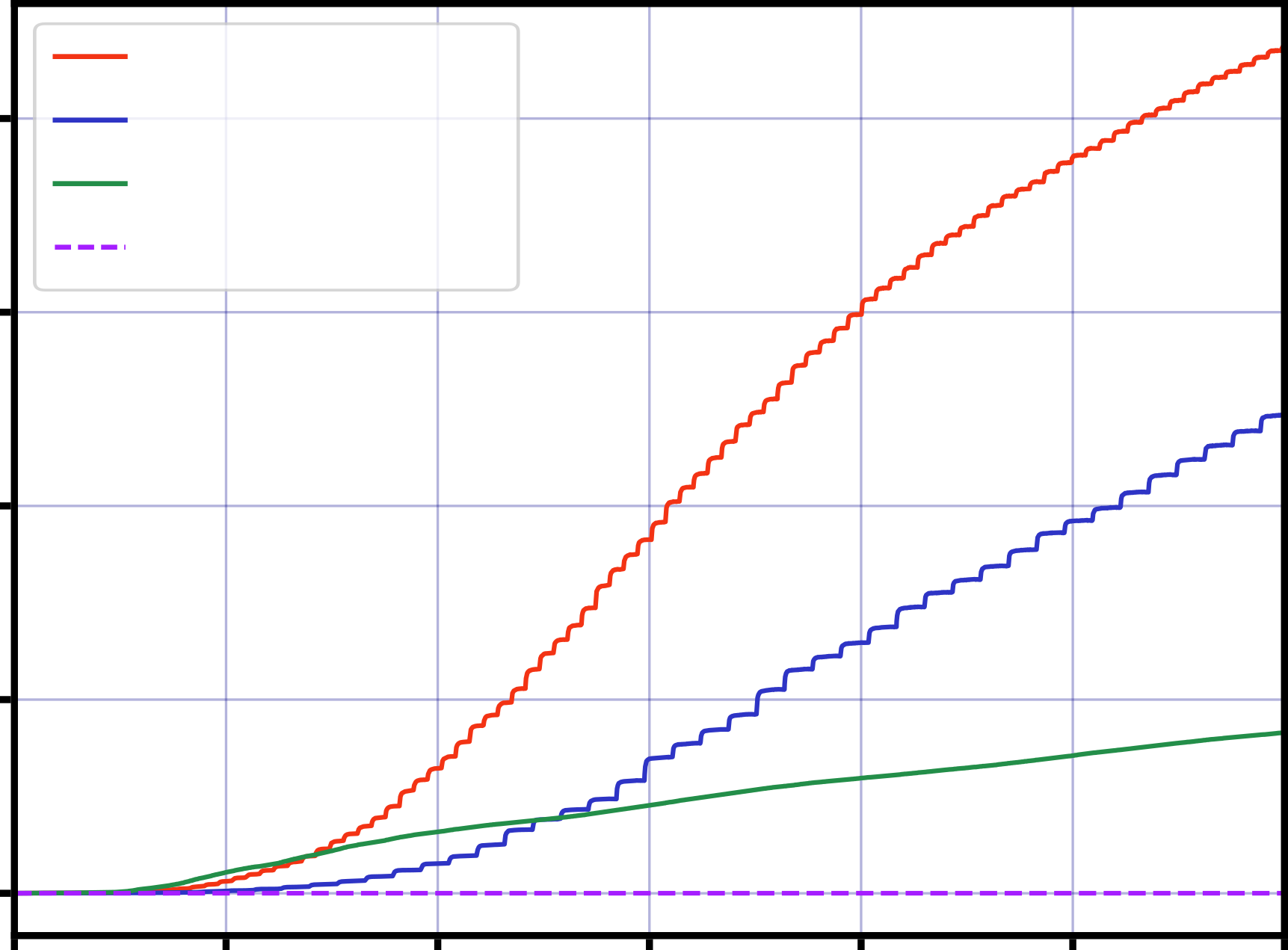}};
    \def\y{-3.25}
    \def\xstart{-4}
    \def\xstep{1.35}
    \node[anchor=center] at (0,-3.8) {t (ms)};
    \node[anchor=center] at (\xstart + 0*\xstep,\y) {0};
    \node[anchor=center] at (\xstart + 1*\xstep,\y) {500};
    \node[anchor=center] at (\xstart + 2*\xstep,\y) {1000};
    \node[anchor=center] at (\xstart + 3*\xstep,\y) {1500};
    \node[anchor=center] at (\xstart + 4*\xstep,\y) {2000};
    \node[anchor=center] at (\xstart + 5*\xstep,\y) {2500};
    \node[anchor=center] at (\xstart + 6*\xstep,\y) {3000};

    \node[anchor=center, rotate=90] at (-5.5,0) {$E_{kin}$ (µK)};
    \def\x{-4}
    \def\ystart{-2.7}
    \def\ystep{1.25}
    \node[anchor=east] at (\x, \ystart + 0*\ystep) {0.00};
    \node[anchor=east] at (\x, \ystart + 1*\ystep) {0.02};
    \node[anchor=east] at (\x, \ystart + 2*\ystep) {0.04};
    \node[anchor=east] at (\x, \ystart + 3*\ystep) {0.06};
    \node[anchor=east] at (\x, \ystart + 4*\ystep) {0.08};

    \def\x{-3.3}
    \def\ystart{2.68}
    \def\ystep{0.425}
    \node[anchor=west] at (\x, \ystart - 0*\ystep) {$2\pi\times \SI{0.8}{\hertz}$};
    \node[anchor=west] at (\x, \ystart - 1*\ystep) {$2\pi\times \SI{1.6}{\hertz}$};
    \node[anchor=west] at (\x, \ystart - 2*\ystep) {$2\pi\times \SI{8.0}{\hertz}$};
    \node[anchor=west] at (\x, \ystart - 3*\ystep) {$2\pi\times \SI{31.8}{\hertz}$};
    \end{tikzpicture}
  \caption{Kinetic energies of simulated time evolution for varying painting frequencies $\omega_{pl}$. The ground state shown in figure~\ref{fig:2D_densities} is propagated while the atoms see the dynamically changing potential of the moving painting beam. We define the frequency $\omega_{pl}$ as the circle frequency for one round trip of the laser beam. A single Gaussian potential of the laser is moved in discrete time steps along the circumference of the ring. This introduces energy into the system and leads to the discrete energy increases of the atoms, which are more prominent for lower frequencies. Initially for low frequencies at $\omega_{pl}= 2\pi\times \SI{0.8}{\hertz}$ (blue) and $\omega_{pl}=2\pi\times \SI{1.6}{\hertz}$ (red), the movement of the painting beam is followed by the atoms resulting in a "stirring" of the cloud at the same frequency. The higher frequency $\omega_{pl}=2\pi\times \SI{7.95}{\hertz}$ (green) however shows how the movement of the painting becomes too fast for the atoms to react to and for $\omega_{pl}=2\pi\times \SI{31.8}{\hertz}$ (purple dashed) the potential becomes effectively constant for the cloud.           
 }\label{fig:2D_energies}
\end{figure}

In the following we show the validity of the time-averaged approach used in the previous sections.
In red-detuned traps, the criterion for time-averaged traps requires the painting frequency to be much larger than the trapping frequency $\omega_{pl} \gg \omega_{trap}$~\cite{Roy2016}. 
However, the concept of trapping frequencies only applies for harmonic potentials and is not applicable here.
We therefore simulate the system described in section~\ref{sec:StaticPotential} and resolve the interactions between the atoms and the potentials in real time. 
The results are shown in figure~\ref{fig:2D_energies}.
We use the kinetic energy of the trapped cloud as an indicator to assess the distortions introduced by the painting beam.
The ground state of a static trap is propagated in time for different painting frequencies $\omega_{pl}$. 
We find that painting frequencies higher than $\omega_{pl}=2\pi\times \SI{31.8}{\hertz}$ for a diameter of $d_{box} = \SI{241.9}{\micro\meter}$ does not increase in the kinetic energy.
Thus, we conclude that the potential changes on a faster time-scale than the atoms can react to. 

Expanding to larger traps with a diameter $d_{bigbox}=\SI{2775.66}{\micro\meter}$ (see figure \ref{fig:2D_densities_bigbox}) the required minimal painting frequency also needs to be increased to $\omega_{pl,min} = 2\pi\times \SI{730.4}{\hertz}$ to sufficiently mimic a static trap, which is well within the experimental capabilities.

\section{Discussion}\label{sec:Discussion}
We have presented a compact and robust apparatus for the creation of large blue-detuned time-averaged optical potentials for the application in microgravity. 
Together with the monitoring capabilities the setup is suitable for the employment in remote and harsh environments.

We performed simulations showing the feasibility of the setup to trap ultracold atomic gases with uniform densities.
We have shown that light powers of a few milliwatts will generate traps with negligible atom loss and that for every trap size a painting threshold can be found to justify a time-averaged approach. 

For the largest trap size, we find that the trap can be described by a power law $V(r)\sim r^p$ with an exponent $p=152$. 
This breaks the threshold of $p \gtrsim 100$, where the thermodynamics of the system is close to an ideal box-like system~\cite{Navon2015,Zobay2005}.
The reported sharpness needs confirmation by performing experiments with ultracold atoms, e.g. by determination of the scaling of critical particle number with temperature~\cite{Navon2021}. 
The confirmation might not be feasible on ground for the large trap sizes as discussed in section~\ref{sec:Introduction} and is instead verified by the numerical simulations in section ~\ref{sec:theory}.

In the future, we are planning studies on the optimal loading of a time-averaged trap, be it with thermal atoms or already condensed atoms. 
While it is in principle possible to simulate the loading into the trap, the required parameters are multitudinous and the simulation is best developed in unison with the experiment. 

Possible applications of our setup range from extremely rarified gases, benefiting the creation of short-lived strongly correlated systems~\cite{Stamper-Kurn2023}, such as Efimov trimers~\cite{Naidon2017}, to mixtures of quantum gases~\cite{Pichery2023} and fundamental studies of the nature of the BEC transition. 
Quantum gas experiments profit from increased atom numbers~\cite{herbst2022}, usually implying large trapping volumes, especially considering magnetically untrappable species, such as ytterbium and strontium.
The large addressable area of our AOD setup enables the realization of multiple independent or interconnected traps.
This allows for the implementation of cold atom colliders~\cite{Rakonjac2012, Thomas2018} and could increase the area of guided atom interferometers\cite{Wu2007}.
It will also benefit research in coherent matter wave optics, strongly interacting gases and molecules and quantum information.
The setup presented here is intended to be deployed onboard the International Space Station as a part of the Bose-Einstein Condensate and Cold Atom Laboratory~\cite{Frye2021}.

\section{Methods}

\subsection{2D interaction strength reduction for box-like BEC}\label{sec:Reduction}
In this sections we show the reduction of the 3D system to an effective 2D system (modified from ~\cite{Boegel2021}), by making assumptions for the shape of the density and therefore the atom-atom interactions of the reduced direction. 
This reduction significantly cuts down on the simulation time while remaining sufficiently accurate for our case study. 
For the 3D system in figure~\ref{fig:reduction_comparison} the reduction lowers the calculation time from one week to only a couple of hours. 
We also compare the weak to the strong interaction regime and find that the reduction works best assuming weak interactions.

We begin by solving the GPE 

\begin{equation}\label{equ:GPE}
i\hbar\frac{\partial}{\partial t}\Psi(\textbf{r},t) = \left[-\frac{\hbar^2}{2m} \frac{\partial^2}{\partial \textbf{r}^2}+V(\textbf{r},t)+gN|\Psi(\textbf{r},t)|^2\right]\Psi(\textbf{r},t)
\end{equation}
in three dimensions $\textbf{r}=(x,y,z)$. 
Here, $N$ is the number of atoms, $g=4\pi \hbar^2 a_s/m$ is the 3D interaction strength per atom with $a_s$ being the s-wave scattering length and $m$ the mass. $V(\textbf{r},t) = V_{\bot}(x,y,t) + V_z(z,t)$ is the total potential, comprised of the painted ring potential $V_{\bot}(x,y,t)$ and two light sheets in $V_z(z,t)$, limiting the extent of the trap in z-direction.

Assuming a pancake shaped cloud with dimensions $L_z \ll L_{x,y}$ ($L_{x,y,z}$ extents in x-, y- and z-direction respectively), the respective trap frequencies are $\omega_z \gg \omega_{x,y}$. 
Therefore, dynamics in the z-direction occur at a comparatively much shorter time scale, making them effectively time-independent from the frame of the x- and y-direction. 
Thus, we perform a separation Ansatz

\begin{equation}\label{equ:separationAnsatz}
\Psi(\textbf{r},t)=\Phi_0(z)\varphi(x,y,t) e^{-\frac{i}{\hbar}\epsilon_0t} 
\end{equation}
where $\Phi_0(z)$ is the now time-independent ground state in z-direction and $\varphi(x,y,t)$ the remaining 2D time-dependent function in radial direction. 
We have defined the average energy along the z-direction as
\begin{equation}\label{equ:epsilon}
\epsilon_0 = \int dz \Phi_0(z)\left(-\frac{\hbar^2}{2m}\frac{\partial^2}{\partial z^2}+V_z(z)\right)\Phi_0(z).
\end{equation}
Here, $\Phi_0$ is normalized by 
\begin{equation}
\int dz \Phi_0^2(z)=1.
\end{equation}

Inserting the Ansatz \ref{equ:separationAnsatz} into equation~\ref{equ:GPE} and integrating in direction z yields the effective 2D GPE: 
\begin{equation}\label{equ:2DGPE}
i\hbar \frac{\partial \varphi(x,y,t)}{\partial t} = \left[-\frac{\hbar^2}{2m} \left(\frac{\partial^2}{\partial x^2}+\frac{\partial^2}{\partial y^2}\right) + V_{\bot}(x,y,t) + \tilde{g}|\varphi(x,y,t)|^2\right]\varphi(x,y,t) ,
\end{equation}
where we have have the reduced effective 2D interaction strength 
\begin{equation}
\tilde{g} = gN\int dz \Phi_0(z)^4 \label{equ:reductionIntegral}
\end{equation}

that depends on the shape of $\Phi_0(z)$, which according to equation~\ref{equ:epsilon} is given by the potential $V_z(z)$. 
In the following we will look at two different interaction strength regimes which impact the choice we make for $\Phi_0(z)$.

\subsubsection{Weak interaction regime}\label{subsec:smallintact}
For low densities the interaction term in equation~\ref{equ:GPE} becomes negligible and the BEC will behave like the classical standing wave in a box potential.
The system can be described by simple standing waves with $L_z$ chosen to fulfill the boundary conditions
\begin{equation}
    \psi_n(z)=\sqrt{\frac{2}{L_z}}\sin\left(\frac{n\pi}{L_z}z\right), n=1,2,3,...
\end{equation}
Choosing the lowest energy level $n=1$ for the ground state and again inserting into \ref{equ:reductionIntegral} we get
\begin{equation}
    \int dz \Phi_0^4 = \int_0^{L_z}  dz \psi_1^4 = \frac{3}{2L_z}
\end{equation}
and finally for the effective 2D interaction strength for low interactions
\begin{equation}
    \tilde{g}=gN\frac{3}{2L_z}.
\end{equation}

\subsubsection{Strong interaction regime}\label{subsec:largeintact}
For a sufficiently large density the BEC will expand to fill out the box-shaped potential confines and will itself also become a box.

\begin{equation}
\psi(z)=
\begin{cases}
    0 & \text{if } z<0 \\
    \frac{1}{\sqrt{L_z}} & \text{if } 0\leq z \leq L_z \\
    0 & \text{if } z <L_z
\end{cases}
\end{equation}

Insert into \ref{equ:reductionIntegral}
\begin{equation}
    \int dz \Phi_0^4(z) = \int_0^{L_z}  dz \psi^4(z) = \frac{1}{L_z}
\end{equation}
And finally for an effective 2D interaction strength for strong interactions we get
\begin{equation}
    \tilde{g}=gN\frac{1}{L_z}
\end{equation}

\subsubsection{Accuracy of effective 2D assumptions}

\begin{figure}[ht]
\centering
\begin{tikzpicture}[>=stealth]
	\node[inner sep=0pt] (Picture) at (0,0)
{\includegraphics[width=390pt]{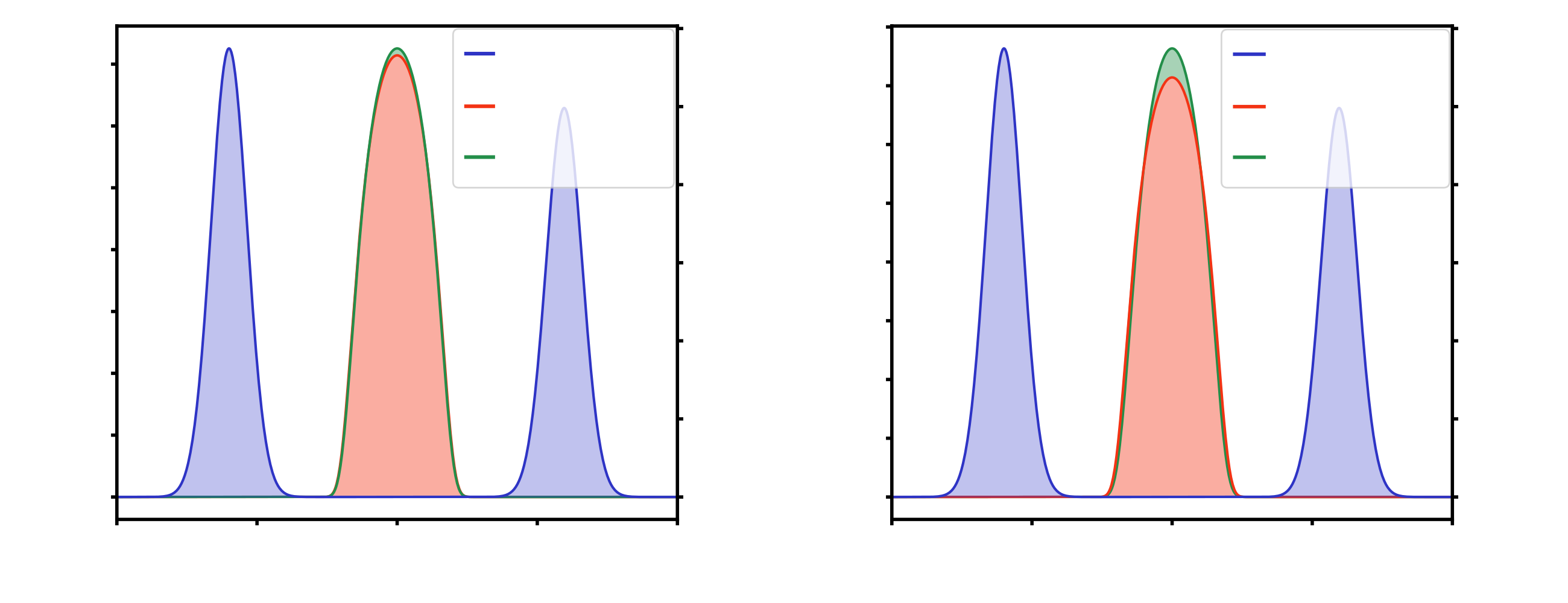}};

    \node[anchor=center] at (-5.45, 2.05) {\textbf{a}};
    \node[anchor=center] at (1.3, 2.05) {\textbf{b}};
    
    \node[anchor=center] at (-3.45,-2.5) {x (µm)};
    \def\y{-2.1}
    \def\xstart{-5.71}
    \def\xstep{1.18}
    \node[anchor=center] at (\xstart + 0*\xstep,\y) {-200};
    \node[anchor=center] at (\xstart + 1*\xstep,\y) {-100};
    \node[anchor=center] at (\xstart + 2*\xstep,\y) {0};
    \node[anchor=center] at (\xstart + 3*\xstep,\y) {100};
    \node[anchor=center] at (\xstart + 4*\xstep,\y) {200};

    \node[anchor=center, rotate=90, align=center, color=red] at (-6.8,0.2) {integrated densities \\ ($10^{-3}$/µm)};
    \def\x{-5.8}
    \def\ystart{-1.65}
    \def\ystep{1.08}
    \node[anchor=east,color=red] at (\x, \ystart + 0*\ystep) {0};
    \node[anchor=east,color=red] at (\x, \ystart + 1*\ystep) {5};
    \node[anchor=east,color=red] at (\x, \ystart + 2*\ystep) {10};
    \node[anchor=east,color=red] at (\x, \ystart + 3*\ystep) {15};

    \def\x{-.2}
    \def\ystart{-1.65}
    \def\ystep{1.35}
    \node[anchor=east, color=blue] at (\x, \ystart + 0*\ystep) {0.0};
    \node[anchor=east, color=blue] at (\x, \ystart + 1*\ystep) {0.1};
    \node[anchor=east, color=blue] at (\x, \ystart + 2*\ystep) {0.2};
    \node[anchor=east, color=blue] at (\x, \ystart + 3*\ystep) {0.3};

    \node[anchor=center] at (2.9,-2.5) {x (µm)};
    \def\y{-2.1}
    \def\xstart{1.08}
    \def\xstep{1.18}
    \node[anchor=center] at (\xstart + 0*\xstep,\y) {-200};
    \node[anchor=center] at (\xstart + 1*\xstep,\y) {-100};
    \node[anchor=center] at (\xstart + 2*\xstep,\y) {0};
    \node[anchor=center] at (\xstart + 3*\xstep,\y) {100};
    \node[anchor=center] at (\xstart + 4*\xstep,\y) {200};

    \def\x{1}
    \def\ystart{-1.65}
    \def\ystep{1.0}
    \node[anchor=east,color=red] at (\x, \ystart + 0*\ystep) {0};
    \node[anchor=east,color=red] at (\x, \ystart + 1*\ystep) {5};
    \node[anchor=east,color=red] at (\x, \ystart + 2*\ystep) {10};
    \node[anchor=east,color=red] at (\x, \ystart + 3*\ystep) {15};
    \node[anchor=east,color=red] at (\x, \ystart + 4*\ystep) {20};

    \node[anchor=center, rotate=90, color=blue] at (6.75,0.3) {$\mathrm{E_{pot}}$ (µK)};
    \def\x{6.6}
    \def\ystart{-1.65}
    \def\ystep{1.35}
    \node[anchor=east, color=blue] at (\x, \ystart + 0*\ystep) {0.0};
    \node[anchor=east, color=blue] at (\x, \ystart + 1*\ystep) {0.1};
    \node[anchor=east, color=blue] at (\x, \ystart + 2*\ystep) {0.2};
    \node[anchor=east, color=blue] at (\x, \ystart + 3*\ystep) {0.3};

    \def\x{-2.52}
    \def\ystart{2.15}
    \def\ystep{0.48}
    \node[anchor=west] at (\x, \ystart - 0*\ystep) {\small{$\mathrm{E_{pot}}$}};
    \node[anchor=west] at (\x, \ystart - 1*\ystep) {\small{$|\Psi_{0,3D}|^2$}};
    \node[anchor=west] at (\x, \ystart - 2*\ystep) {\small{$|\Psi_{0,2D}|^2$}};

    \def\x{4.3}
    \def\ystart{2.15}
    \def\ystep{0.48}
    \node[anchor=west] at (\x, \ystart - 0*\ystep) {\small{$\mathrm{E_{pot}}$}};
    \node[anchor=west] at (\x, \ystart - 1*\ystep) {\small{$|\Psi_{0,3D}|^2$}};
    \node[anchor=west] at (\x, \ystart - 2*\ystep) {\small{$|\Psi_{0,2D}|^2$}};
\end{tikzpicture}
  \caption{Comparison of 2D and 3D ground state solutions, to check the validity of the interaction strength reduction. Here it is checked for $N=7500$ atoms, a box diameter of $d_{box}=\SI{240.19}{\micro\meter}$ and a z-extent $L_z=\SI{20}{\micro\meter}$. \textbf{a} The ground states using the effective interaction strength $\tilde{g}$ of the \textit{low interaction regime} (\ref{subsec:smallintact}). Both the ground states $|\varphi_{0,2D}|^2$ (green) (Eq. \ref{equ:2DGPE}) and the $|\Psi_{0, 3D}|^2$ (red) (Eq. \ref{equ:GPE}) are integrated over their y- and (y, z)-direction respectively to be comparable, whereas the potential energy $E_{pot}$ (blue) is a 1D slice at $y=\SI{0}{\micro\meter}$ for reference. \textbf{b} The solution for the same system, but utilizing the effective interaction strength $\tilde{g}$ of the \textit{strong interaction regime} (\ref{subsec:largeintact}). For this specific setup, the low interaction regime shown in a) gives a much closer match between the 2D and 3D densities than for the strong interaction regime shown in b).  
 }\label{fig:reduction_comparison}
\end{figure}
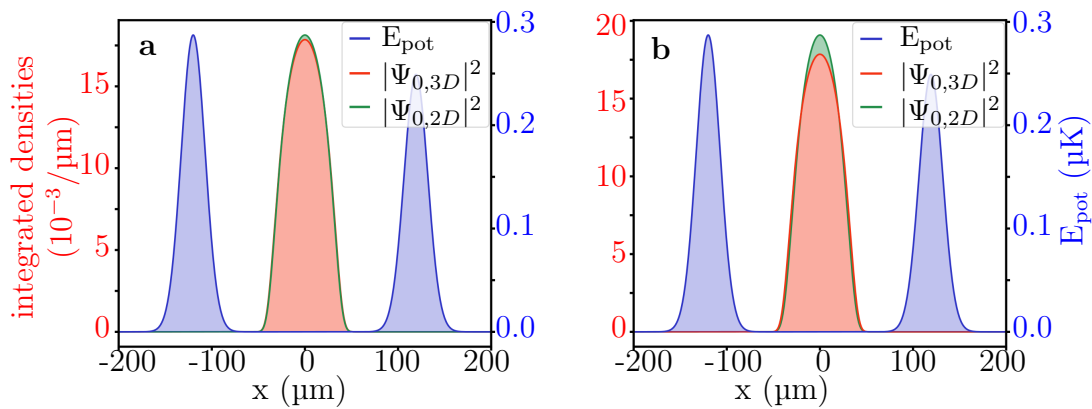

The large and weak interaction regimes are compared to the full 3D simulation with identical parameters. 
For the weak interaction regime, the difference between the 3D $|\Psi_{0,3D}|^2$ (red) and the effective 2D $|\Psi_{0,2D}|^2$ (green) densities is negligible (figure~\ref{fig:reduction_comparison}(a)). 
We also observe a negligible difference when simulating the dynamics. 
The difference is however much more noticeable for the strong interaction regime shown in figure~\ref{fig:reduction_comparison}(b).
In this publication, we therefore limit ourselves to systems in the weak interaction regime.

\subsection{Measurement setup}\label{sec:measurement}
For the characterisation, a temperature stabilised, self-build external cavity diode laser~\cite{Baillard2006} provides light at a wavelength of \SI{764}{\nano \meter}.
The laser is neither intensity nor frequency stabilized.
We take images with a Grasshopper GS3-U3-15S5M-C from \textit{Teledyne FLIR LLC}.
By removing its protection window we avoid interference fringes.
Unless stated otherwise, we synchronise the exposure time of the camera with the time to create a single realization of an optical potential shape. 
We confirmed this by increasing an initially short exposure time step-wise until a completed pattern is observed.

\subsection{Smoothness of the potentials}\label{sec:smooth}
The confining potential for the atoms should feature low intensity variations to avoid density modulations or leakage.
In this section we show the different steps to minimize intensity variations of the optical potentials.

First, we linearize the deflection efficiency $\varepsilon(f)$ of the AOD, which is typically not uniform in frequency. 
We measure $\varepsilon(f)$ by scanning the driving frequencies for each axis separately in the range of $f_{mid}-\SI{15}{\mega\hertz}$ to $f_{mid}+\SI{15}{\mega\hertz}$, with $f_{mid}$ being the central drive frequency ($\approx \SI{100}{\mega\hertz}$), while recording the deflected beam on a photo diode.
We then fit a polynomial to $\varepsilon(f)$ per axis to flatten the optical power output of the AOD by adjusting the RF power.

\begin{figure}[ht]
\centering
	  \begin{tikzpicture}[>=stealth]
	\node[inner sep=0pt] (Picture) at (0,0)
	{\includegraphics[width=429pt]{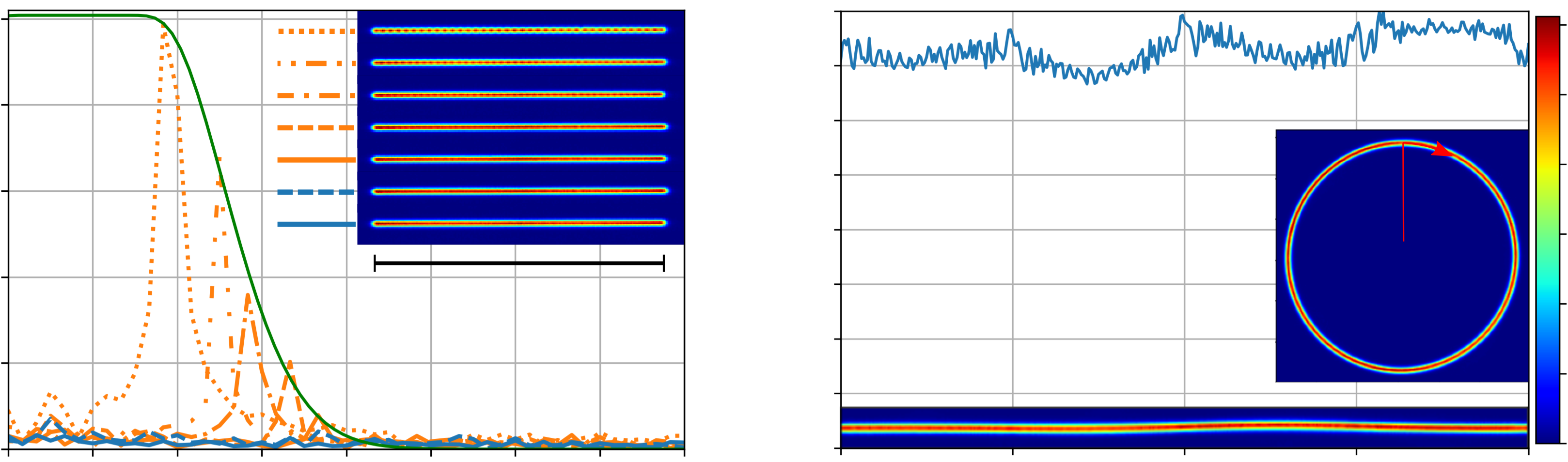}};

	
	\node[anchor=west] at (-8.5,2.5) {\large \textbf{a}};
	\node[anchor=west] at (0.5,2.5) {\large \textbf{b}};

	\node[rotate=90, align=center] at (-8.75, 0) {normalized \\ oscillation amplitude};
	
	\def\spacingRx{0.83}
	\def\Lupperleft{2.1}
	\def\rgr{-7.4}
	\node[anchor=east] at (\rgr,\Lupperleft-0*\spacingRx) {1};
	\node[anchor=east] at (\rgr,\Lupperleft-1*\spacingRx) {0.8};
	\node[anchor=east] at (\rgr,\Lupperleft-2*\spacingRx) {0.6};
	\node[anchor=east] at (\rgr,\Lupperleft-3*\spacingRx) {0.4};
	\node[anchor=east] at (\rgr,\Lupperleft-4*\spacingRx) {0.2};
	\node[anchor=east] at (\rgr,\Lupperleft-5*\spacingRx) {0};
	
	\node[anchor=west] at (-6,-2.75) {spatial frequency (1/mm)};
	\def\bgr{-2.35}
	\def\spacingLx{0.815}
    \def\Llowerleft{-7.8}	
	\node[anchor=west] at (\Llowerleft-0*\spacingLx,\bgr) {20};
	\node[anchor=west] at (\Llowerleft+1*\spacingLx,\bgr) {25};
	\node[anchor=west] at (\Llowerleft+2*\spacingLx,\bgr) {30};
	\node[anchor=west] at (\Llowerleft+3*\spacingLx,\bgr) {35};
	\node[anchor=west] at (\Llowerleft+4*\spacingLx,\bgr) {40};
	\node[anchor=west] at (\Llowerleft+5*\spacingLx,\bgr) {45};
	\node[anchor=west] at (\Llowerleft+6*\spacingLx,\bgr) {50};
	\node[anchor=west] at (\Llowerleft+7*\spacingLx,\bgr) {55};
	\node[anchor=west] at (\Llowerleft+8*\spacingLx,\bgr) {60};

	\node[anchor=west] at (3.1,-2.7) {angle (°)};
	
    \def\spacingX{1.57}
    \def\Rlowerleft{0.35}
	\node[anchor=west] at (\Rlowerleft-0*\spacingX,\bgr) {0};
	\node[anchor=west] at (\Rlowerleft+1*\spacingX,\bgr) {90};	
	\node[anchor=west] at (\Rlowerleft+2*\spacingX,\bgr) {180};
	\node[anchor=west] at (\Rlowerleft+3*\spacingX,\bgr) {270};
	\node[anchor=west] at (\Rlowerleft+4*\spacingX,\bgr) {360};
	
	\def\Rupperleft{2.1}
    \def\spacingY{0.517}
    \def\lgr{0.55}
	\node[anchor=east] at (\lgr,\Rupperleft-0*\spacingY) {1.0};
	\node[anchor=east] at (\lgr,\Rupperleft-1*\spacingY) {0.9};
	\node[anchor=east] at (\lgr,\Rupperleft-2*\spacingY) {0.8};
	\node[anchor=east] at (\lgr,\Rupperleft-3*\spacingY) {0.7};
	\node[anchor=east] at (\lgr,\Rupperleft-4*\spacingY) {0.6};
	\node[anchor=east] at (\lgr,\Rupperleft-5*\spacingY) {0.5};
	\node[anchor=east] at (\lgr,\Rupperleft-6*\spacingY) {0.4};
	\node[anchor=east] at (\lgr,\Rupperleft-7*\spacingY) {0.3};
	\node[anchor=east] at (\lgr,\Rupperleft-8*\spacingY) {0.2};
	
	\node[rotate=90] at (-0.35, 0) {norm. Intensity};
	
	\node[anchor=center] at (-2.4,-0.5) {\SI{1.13}{\milli\meter}};
	
	\node[anchor=west] at (7.4, -2) {0};
	\node[anchor=west] at (7.4, 2) {$I_{max}$};
	\end{tikzpicture}

  \caption{Analysis of time-averaged optical potentials. \textbf{a} Fragmentation of painted lines, depicted in the inset. The spacing between beam spots is decreased from top to bottom. Fourier transformations along the lines reveal the existence (orange lines) or absence (blue lines) of dominant frequency components. The amplitude of the oscillation is normalized such that far separated beams give an oscillation amplitude of 1. The green curve fits the maxima of the orange curves (see section~\ref{sec:Fragmentation}). \textbf{b} Analysis of a painted circle (inset on the right). The blue graph shows intensity across the circle's circumference. At the bottom we show an unfolded image of the circle. The inset marks the \SI{0}{\degree} angle and the direction of beam movement.}\label{fig:PotAnalysis}
\end{figure}
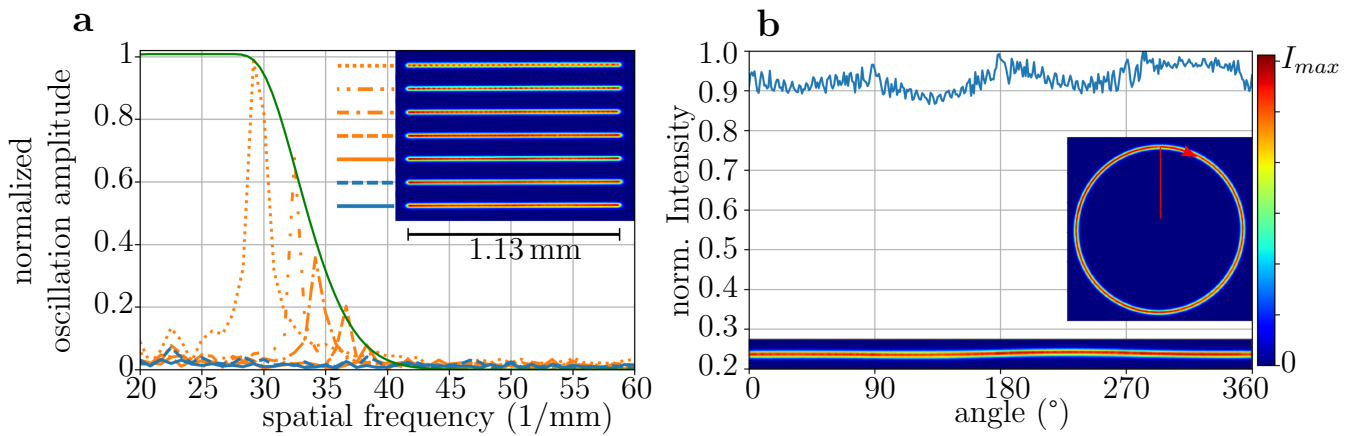

Next, we investigate the spacing of consecutive beam positions.
A small spacing $s$ will result in a smooth optical potential, but a large spacing increases the repetition rate of the potential.
We draw several lines with a constant length, each with a certain, fixed spacing of individual points (inset of figure~\ref{fig:PotAnalysis}(a)). 
A Fourier analysis of the measured intensity (figure~\ref{fig:PotAnalysis}(a)) is performed, where resulting smooth lines correspond to a Fourier spectrum revealing only low-amplitude noise (blue lines).
When the separation of the points becomes visible, a significant frequency component appears in the Fourier spectrum (orange lines).
These frequency components originate from fragmentation of the lines.
The green line shows a fit of our theoretical model (see section~\ref{sec:Fragmentation}) to the frequencies and the amplitudes. 
We find that the method provides a robust and unambiguous way to identify an optimal spacing, more robust than a direct analysis of the intensity, which can be overlayed by noise.
At a spatial frequency of \SI{42}{\per\milli\meter} the fitted model drops below the noise background (blue curves), which corresponds to a spacing over beam radius (1/e$^2$) of $1.19$.
Using the measured beam waist of $w_0 = \SI{19}{\micro\meter}$ a spacing of the individual beams lesser than \SI{22.6}{\micro\meter} is below our detection threshold.
However, in section~\ref{sec:theory} we used a spacing of \SI{40}{\micro\meter} to create homogeneous BECs, showing that the atoms are less sensitive to an uneven trapping potential than a CCD camera, most likely due to the atoms only experiencing the potential close to the trap bottom.

Finally, we investigate the ability to create circular traps which are of special interest because they provide a high degree of symmetry.
In figure~\ref{fig:PotAnalysis}(b), we show that the intensity variations along a circle are less than \SI{10}{\percent}.
The bottom inset shows a linear projection of the circle around the center.
The circle is slightly elliptical, with deviations from a perfect circle being below $\SI{10}{\micro\meter}$ for a circle with diameter $d_{box}> \SI{1}{\milli\meter}$.
We attribute the non-zero ellipticities to the slightly angled passage of the laser beam through the AOD, which was optimized to the highest deflection efficiency. 
However, the angle results in different interaction times with the sound wave across the beam, effectively realizing a multiplex operation AOD and distorting the circle.

To gauge the impact of the intensity variations, we have performed simulations similar to the ones shown in figure~\ref{fig:2D_energies}, where we additionally account for power fluctuations of the painting laser by introducing a spatial sine modulation of the Gaussian beam power during the simulations.
This way each Gaussian is painted with a different power along the circle. 
We compare the trap with varying optical powers to the non-modulated case. 
We find for a \SI{10}{\percent} modulation of the beam power that the effect on the kinetic energy for a painting frequency $\omega_{pl}=2\pi\times \SI{31.8}{\hertz}$ is on the order of $10^{-7} \mu K$.

\subsubsection{Fit function for figure~\ref{fig:PotAnalysis}(a)}\label{sec:Fragmentation}
The spacing $s$ of the beam spots determines the depth of the fragmentation $\Delta$ -- the smaller the spacing, the lower the fragmentation. 
To derive $\Delta$ we model the beam spots as Gaussian functions.
We define the function $f_k$ as the sum of $2k+1$ Gaussian beams

\begin{equation}
    f_k(x) = \sum_{n=-k}^k A\cdot \exp\left(\frac{-(x-ns)^2}{w^2}\right),
\end{equation}
with $A$ being the amplitude and $w$ the width of the beam. 
The difference in the amplitude of the central beam when considering an infinite number of beams versus only five beams on each side is approximately $1e-5$.
This justifies the approximation of using an infinite number of Gaussian beams with equal spacing $s$ which we adopt for our study.
The fragmentation depth $\Delta$ will be the difference of $f_\infty(x=0)$ and $f_\infty(x=-s/2)$, e.g. 
\begin{equation}\label{equ:infinite_sum}
    \Delta(a) = A \cdot \sum_{n =-\infty }^{\infty} \exp\left(-a^2n^2\right)-  A \cdot\sum_{n =-\infty }^{\infty} \exp\left(-a^2 (n+1/2)^2\right).
\end{equation}
We have defined the dimensionless parameter $a=s/w$. 
The infinite sums in equation \ref{equ:infinite_sum} can be written in terms of Jacobi theta functions, which are defined as
\begin{eqnarray}
    \theta_2(q) &= \sum_{n =-\infty }^{\infty} q^{(n+1/2)^2} \label{equ:theta2} \\
    \theta_3(q) &= \sum_{n =-\infty }^{\infty} q^{n^2} \label{equ:theta3}.
\end{eqnarray}
Comparing equation \ref{equ:infinite_sum} with equations \ref{equ:theta2} and \ref{equ:theta3} yields 

\begin{equation}\label{equ:FitFct}
    \Delta(a) = A\cdot\left[\theta_3\left(e^{-a^2}\right) - \theta_2\left(e^{-a^2}\right)\right].
\end{equation}

Rewriting equation \ref{equ:infinite_sum} to a sum of Jacobi theta functions simplifies the evaluation of measurement data since they are part of standard software packages. 
For the analysis performed in this paper, we used \textit{Python}, with the package \textit{mpmath} containing the function \textit{mpmath.jtheta()}~\cite{jtheta}.
To perform the fitting routine for figure~\ref{fig:PotAnalysis}(a), one has to transform $a \rightarrow 1/a$, because the x-axis is showing the modulation frequency $f$, which is proportional to $1/s$.

Equation~\ref{equ:FitFct} only has the amplitude $A$ as a parameter for optimization.
We introduce a shift $\mu$ and a scaling factor $d$ to have the fitting routines converge. Summing up, we fit the following function:

\begin{equation}
    \Delta(f) = A\cdot\left[\theta_3\left(e^{-(d/(f-\mu))^2}\right) - \theta_2\left(e^{-(d/(f-\mu))^2}\right)\right].
\end{equation}

\subsection{Determination of the best fit exponent of a power law potential}\label{sec:AppendixExponent}
Here, we describe how to obtain the best fit exponent to a potential of the form $V(r) \sim r^p$.
\begin{enumerate}
    \item To interpolate sparse measurement data, we fit a double Gaussian function of the form $y'(x) = A_1 \exp(-2\cdot(x-\mu_1)^2/w^2) + A_2 \exp(-2\cdot(x-\mu_2)^2/w^2)$, where $A_1$, $A_2$, $\mu_1$, $\mu_2$, and $w$ are the parameters to be optimized. This fitting process aligns $y'(x)$ with the experimentally measured data $y(x)$, represented as blue dots in figure~\ref{fig:LargeBox}(b). This enhances the data density by two orders of magnitude, mitigating the fact, that very few data points in figure~\ref{fig:LargeBox}(b) are significantly above zero.
    \item The enhanced data set is used for conventional fitting routines (e.g. the Levenberg-Marquardt algorithm) with fixed exponent $p$.  The fit function $y''(x) = a \cdot x^p$ contains only $a$ as the single free fit parameter.
    \item For every fixed exponent $p$ we compute a squared residual $r = (y(x)-y''(x))^2$ and determine the mean $\bar{r}$.
\end{enumerate}

\noindent{The exponent that gives the lowest mean squared residual $\bar{r}$ is taken as the best fit result.}

\subsection{Driving electronics and control software}\label{Sec:Electronics}
The deflection angle of the beam passing the AOD is controlled by two radio frequency signals for x- and y-direction with up to \SI{2}{\watt} of power per direction.
Typically, these frequencies are generated by bulky, commercial arbitrary waveform generators~\cite{Henderson2009,Bell2016} or software-defined radios~\cite{herbst2022}.
However, we present an alternative approach which is suitable for compact apparatuses. 
We utilize a direct digital synthesizer (DDS) AD9958 chip from \textit{Analog Devices} which creates the frequencies and amplitudes for x and y deflection at the AOD.
The DDS chip runs two numerically controlled oscillators with look-up tables for sine-functions, power factors and 10 bit digital analog converter outputs at \SI{400}{\mega\hertz} (\SI{500}{\mega\hertz} maximum). 
It accepts 32 bit wide frequency control values, whereas the processor is only 16 bits wide in order to reach a clock speed of \SI{100}{\mega\hertz}. 
The DDS is controlled by an \textit{Intel 10M08} field-programmable gate array (FPGA).
The FPGA runs a Forth compiler (Mecrisp-Ice~\cite{Mecrisp2022}), which uses only a few kilobytes of memory and runs on a custom single cycle per instruction stack processor, to define the radio frequencies, and therefore the optical potentials.

To implement the potential shapes and animations, experiment operator describes them with simple statements in Forth source codes.
It is possible to create lines, circles (using the Minsky algorithm~\cite{Neal1990}), boxes and single points using predefined functions.
The shapes are calculated on-the-fly, which enables parametric changes of the shapes over time, allowing the generation of animations.
Also, look-up tables are possible, which we use to create the dog in the center bottom picture of figure~\ref{fig:Potentials_exmaples}.

To simplify programming for the end-user, a first-in-first-out (FIFO) buffer is inserted between the processor and the logic driving the synthesis chip. 
The data from the FIFO is subsequently used to update the frequency synthesis at a fixed rate, with the only timing requirement imposed on the user generated code to calculate pixel data quickly enough to avoid buffer underrun. 
It is possible to intertwine a lot of quick calculations with occasional longer computations as long as the buffer keeps getting filled. 
Additionally, buffer-full detection ensures that no pixel data is lost.
A cycle-accurate simulation using Verilator~\cite{Verilator2022} is provided so that scientists can develop the Forth code prior loading into hardware.

Finally, we notice the absence of a position depend blurring of the circular potential in figure~\ref{fig:PotAnalysis}(b), which indicates a negligible phase noise. 
Presence of significant phase noise would lead to different contour widths in figure~\ref{fig:PotAnalysis}(b) around the diagonals in comparison to the main axes due to 
the simultaneous change in frequency in x- and y-direction.

\section{Acknowledgements}
We thank Sven Abend and Brendan Rhyno for reviewing the manuscript.
We acknowledge the work from the BECCAL Science Definition Team, namely D. Stamper-Kurn, H. Müller, N. Lundblad, B.K. Stuhl, D. Becker, W.P. Schleich and M. Krutzik. 
We thank S.-W. Chiow and N. Yu for fruitful discussions. 
Funded by the Deutsche Forschungsgemeinschaft (DFG, German Research Foundation) under Germany’s Excellence Strategy – EXC-2123 QuantumFrontiers – 390837967.

\section{Author contributions}
K. F-A., C.S. and E.R. created the concept of the paper. 
K. F-A. wrote the manuscript with contributions by M. Glay, N.G. and M.K..
The mechanical setup was designed by K. F-A..
K. F-A. carried out the measurements and performed the data evaluation with contributions by M. Glae. under supervision of W.H., C.S. and E.R..
M. Glay and H.A. performed the simulations with contributions by K. F-A. under supervision of N.G and C.S and using a code base developed by S.S..
M. K. developed the control electronics. 
The work is lead by E.R.. 

\section{Competing interests}
We have no competing interests to declare.

\bibliographystyle{vancouver} 
\bibliography{literature}

\newpage
\appendix

\end{document}